\documentclass[journal, onecolumn]{IEEEtran}

\usepackage{lineno,hyperref}
\modulolinenumbers[5]
\usepackage{cite}
\usepackage{amsmath,amssymb,amsfonts}
\usepackage{graphicx}
\usepackage{textcomp}
\usepackage{multirow}
\usepackage{caption}
\usepackage{blindtext}
\usepackage{xcolor}
\usepackage[utf8]{inputenc}
\usepackage[T1]{fontenc} 
\usepackage{mathtools}
\usepackage{tabularx}
\usepackage{subfigure}
\usepackage{dblfloatfix}
\usepackage{fixltx2e}
\usepackage[justification=centering]{caption}

\usepackage{multicol}
\usepackage{lipsum}
\usepackage{mwe}
\usepackage{longtable}
\usepackage[normalem]{ulem}
\usepackage{algorithm,algpseudocode}
\usepackage{varwidth}
\usepackage{amsmath,amssymb,amsfonts}
\usepackage{graphicx}
\usepackage{textcomp}
\usepackage{multirow}
\usepackage{caption}
\usepackage{subcaption}
\usepackage{hyperref} 
\usepackage{blindtext}
\usepackage[utf8]{inputenc}
\usepackage{algorithm,algpseudocode}
\usepackage{varwidth}

\begin{document}

\title{Multi-criteria recommendation systems to foster online grocery}
\author{Manar Mohamed Hafez, Rebeca P. Díaz Redondo, Ana Fernández-Vilas,  Héctor Olivera Pazó
\thanks{Manar Mohamed Hafez (m.mohamed@aast.edu) is with College of Computing and Information Technology, Arab Academy for Science, Technology and Maritime Transport (AASTMT)-Smart Village, P.O. Box 12676, Giza, Egypt}
\thanks{Rebeca P. Díaz Redondo (rebeca@det.uvigo.es), Ana Fernández-Vilas (avilas@det.uvigo.es) and Héctor Olivera Pazó (iclab@det.uvigo.es) are with atlanTTic, Universidade de Vigo; Vigo, 36310, Spain.}}

\maketitle

\abstract{With the exponential increase in information, it has become imperative to design mechanisms that allow users to access what matters to them as quickly as possible. 
The recommendation system ($RS$) with information technology development is the solution, it is an intelligent system. Various types of data can be collected on items of interest to users and presented as recommendations. $RS$ also play a very important role in e-commerce. 
The purpose of recommending a product is to designate the most appropriate designation for a specific product. The major challenges when recommending products are insufficient information about the products and the categories to which they belong. 
In this paper, we transform the product data using two methods of document representation: bag-of-words (BOW) and the neural network-based document combination known as vector-based (Doc2Vec). We propose three-criteria recommendation systems (product, package, and health) for each document representation method to foster online grocery, which depends on product characteristics such as (composition, packaging, nutrition table, allergen, etc.). For our evaluation, we conducted a user and expert survey. Finally, we have compared the performance of these three criteria for each document representation method, discovering that the neural network-based (Doc2Vec) performs better and completely alters the results.}


\section{Introduction}
\label{sec:introduction}

According to \cite{article-digital-transformation}, digital transformation facilitates new ways of value creation at all stages of the consumer decision process: pre-purchase (need recognition, information search, consideration or evaluation of alternatives), the purchase (choice, ordering, payment), and the post-purchase (consumption, use, engagement, service requests). This value creation is especially relevant in retailing to ensure competitiveness and gain a larger market share. Digital transformation came hand in hand with the penetration of mobile devices and data science in e-commerce. Although digital transformation \cite{article-unpacking-digital-tramsformation} has been addressed from several approaches; multi-channel solutions, user modeling, Internet of Things, etc., all of them rely to some extent on the availability of information on operations, supply chains, and consumer and shopper behaviors. One of the imperatives in this digital transformation is obtaining a view of customer insights.

From the early steps (Amazon, 2003~\cite{linden2003amazon}), the time to select the desired product has been the main issue for customers, especially if the high volume and rhythm of incorporation of products are considered. From more than two decades, Recommender Systems (RS) in e-commerce have tried to provide the most suitable products of services, to mitigate the product overload problem and to narrow down the set of choices~\cite{thorat2015survey, grbovic2015commerce, shen2020collaborative, li2020evaluation}. Success of major products \& service providers mainly relies on $RS$, such as Amazon~\cite{linden2003amazon}, Netflix~\cite{bennett2007netflix}, and Google~\cite{das2007google}. $RS$s improve customer satisfaction by reducing customer search efforts and as a consequence, they increase product/service sales.  $RS$s provide users with items based on their interests, the preferences of other users and the item attributes.  The recommendation can be carried out form several approaches depending on the type of data collected and the ways it is used by the $RS$: Content-Based ($CB$) filtering, Collaborative Filtering ($CF$), and hybrid. 
Both systems $CB$ and $CF$ are widely used, and specially the item-based collaborative filtering where the similarity between items is calculated using users' ratings of those items. (developed by Amazon~\cite{linden2003amazon}).

Although $RS$s are used by users regularly in almost all digitalized sectors, its popularization in the grocery market, i.e., a retail store that primarily sells food products, has been delayed as a consequence of the low penetration of online grocery, the implementation of e-commerce for grocery goods. Recently, as well as in other sectors, the grocery industry is harnessing digital to innovate through data-drive business models. Online grocery is considered a central element in the new normal. In this respect, grocery recommendation uses customer's shopping history and product information to address various added value scenarios; predicting customers' future shopping, selecting best value for money products, offering new products user may like, etc. Besides, the availability of data about products and shopping positively affects the retailer by easing a sustainable business; offers \& featured products, stock management, customer profiling, etc. 

To meet the challenges above, in this paper, we use two document representation methods; BOW and Doc2Vec, to manage product data. We also address the three-criteria recommendation systems; Product, Package, and Health for each document representation model to the specific problem of, given a source product $P$, applying $RS$s to suggest similar alternative products where similarity is defined on the basis of a product taxonomy, as well as product characteristics; composition, packaging, nutrition table, allergens, etc. The solution to this problem supports various regular use cases in the grocery market, such as out of stock products, inventory clearance, best value options, new products, etc. In order to obtain the recommender model and to validate them, we use a real grocery dataset, referred to as MDD-DS, provided by \textit{Midiadia}, a Spanish company that works on grocery catalogs. MDD-DS was constructed by analyzing the product's information (product labeling) and by experts' manual annotation so that products are assigned to a specific variety in a hierarchical structure for products. Therefore, the major contributions of this research work are the following ones: 
\begin{enumerate}
    \item Definition of an appropriate data structure to manage the different kinds of information linked to commercial products (especially in the food industry). 

     \item Definition and identification of the appropriate document representation that works with MDD-DS to represent the products.

    \item Design and implementation of a RS that automatically provides alternative products when the user's choice is not available. The RS do not work with user's profile, it is exclusively based on the product's characteristics and the available catalogue.
    \item Design of three recommendation approaches based on the product's characteristics; composition, packaging, nutritional table, allergens, etc. 
    \item Proof of concept and validation to test the RS performance. We have conducted a survey for users and for experts to evaluate the RS approaches.
\end{enumerate}

The rest of this document is organized as follows: In~\autoref{sec:RS}, we briefly reviewed $RS$ and document representation methods to manage product data in $RS$. The grocery MDD-DS is describing in \autoref{sec:dataset}. In~\autoref{sec:Methodology}, the recommendation methodology is introduced with three specific approaches to product similarity, based on product composition, packaging, and healthy characteristics. To implement these three approaches to product similarity, we deployed two kinds of document representation techniques: a simple BOW (Bag of Words, in~\autoref{sec:RSnonML}) and a neural network-based word embedding, Doc2Vec in~\autoref{RS-ml}. For the two product representation models, experimental evaluation and discussion are described in~\autoref{sec:results}. Finally, in~\autoref{sec:conclusions}, we conclude the current work with some future research directions.

\section{Recommender Systems}
\label{sec:RS}

$RS$ are a fundamental task for e-commerce, as the personal $RS$ recommends providing items or products that satisfy the interests of different users according to their different interests and also recommends unknown items for the users that satisfy their interests~\cite{hendrick2018recommendations, kumar2020recommendation}. As mentioned above, the three most commonly used methods in the $RS$ are $CB$ filtering, $CF$, and hybrid approach. 

\textbf{CB} filtering~\cite{zhang2014research,pirasteh2014item,bag2019efficient} is one of the standard techniques used by RS. CB identifies items based on an analysis of the item's content, similar to items known to be of interest to the user. For example, a CB website recommendation service can work by analyzing the user's favorite web pages to generate a profile of commonly occurring terms. Then use this profile to find other web pages that include some or all of these terms.

CB technique has several issues and limitations~\cite{van2000using,lops2011content,saravanan2015design}. For example, (i) having no mechanism to assess the quality of an item supported by CB methods. Furthermore, CB methods generally require items to include some type of content that is amenable to feature extraction algorithms. As a result, CB technique tend to be ill-suited for recommending products, movies, music titles, authors, restaurants, and other types of items with little or no useful and analyzable content; (ii) CB is also have another problem that they rarely reflect current user community preferences. In a technique that recommends products to users, for example, there is no mechanism to favor items that are currently "hot sellers." Moreover, existing systems do not provide a mechanism to recognize that the user can search for a particular type or category.

\textbf{CF}~\cite{schafer2007collaborative, elahi2016survey} is another common recommendation technique. In general, the CF recommends the item to the user based on a community of user interests, without any analysis of the item content. CF idea is to build a personal profile of ratings data through each item sold and rate it through the user. Besides the CF technique's concept to recommend the item to the user, the user's profile is initially compared with other users' profiles to identify one or more similar users. These similar users' highly-rated items are recommended to the user. A significant benefit of CF is that it overcomes the previously mentioned shortcomings of CB filtering.

The main issue in the above is how to measure user similarity. This problem inspires memory-based methods~\cite{yu2004probabilistic}, which can be implemented as user-based~\cite{koohi2016user} or item-based~\cite{sarwar2001item, gao2011userrank}. User and item-based methods have similar mechanisms, but item-based methods are used more to perform better at scale and with a lower rating density. 

A \textbf{hybrid approach} is an approach that combines CB and CF (user-based and item-based) filtration approaches with attempts to eliminate their flaws and provides a more efficient result. It usually perform better than either filtering method alone. 
Here, the hybrid approach does combine between the CB and CF to solve the significant problems that are cold start~\cite{wei2017collaborative} and sparsity problems~\cite{alaa2020personalized}. The cold start problem occurs when there is not enough new user data or ratings for a new item, so it is difficult to make recommendations for that new user or present the new item to a user. Regarding sparsity, it occurs when the user has not rated most of the items and the ratings are sparse.

In our work, we have some issues in providing a recommendation service and associated methods for generating personalized items. Science, the recommendation is based on the user's interests without considering the user profile. 
In this paper focused solely on the user's interest and how to recommend suitable items to each user. The benefit of this work is also that recommended items are identified by lists of similar items to the desired item. As mentioned above, in our paper worked on combine between CB filtering and CF (item-to-item) such as Amazon~\cite{linden2003amazon}.
\textbf{Amazon} has invented an algorithm that began looking at items themselves. It analyzes the recommendations through the items purchased or rated by the user and matches them with similar items, using metrics and composing a list of recommendations. That algorithm is called "item-based collaborative filtering.". This approach was also very appropriate and faster, especially for huge data sets. It was also developed in 2017~\cite{smith2017two} to aggregate data about the user to develop a $RS$ to rely on the data and the user behavior in selecting the items. It is still based only on the analysis of the items. However, it combines the analysis of the items with the user’s data and choices. Regarding the related works, we see that the most widely used in the previous works is collaborative filtering. As shown in the following paragraphs.  

In~\cite{esparza2012mining} used a collaborative filtering method to create the proposal for various items using accessible ratings and comments on Twitter. The authors have also evaluated the reviews given by blipper (a review website) for four unique products using the CF method. 

When dealing with video as data to find suitable items for the user, there are also research works that apply collaborative filtering to recommend products through this kind of data. For instance, in~\cite{jin2010myspace} introduced an approach that includes item-to-item collaborative filtering to discover exciting and meaningful videos among the large-scale videos. This method runs on Qizmt, which is a.NET MapReduce framework. The $RS$ in~\cite{wilkinson2020systems} also depends on monitoring the video content the user watches, the customer carrier database, and the vector database of products; therefore, the idea is to identify an item related to a part of the video content the user viewed that, and consequently determine the product category associated with the item, then analyze the characteristics of items similar to the item. That has been identified through the video's visualization, and it compares the customer value vectors and the product characteristics vectors.
Moreover, start showing the recommended product to the customer. Other approaches take user interactions into account to recommend the right products. For instance, in~\cite{singh2020event}, the recommender system collaborative filtering uses user interactions and keeps them to benefit the recommendation. It does not stop at the items that have been selected only from the users, but the proposed system is related to the category of items. 

Recommendation systems usually require a large amount of user data. Safeguarding the privacy of this information is an important aspect that must be taken into account. For instance, in~\cite{xu2019blockchain}, an arbitrable remote data auditing scheme is proposed. This is based on a non third-party auditor  for the network storage-as-a-service paradigm. The authors have designed a network storage service system based on blockchain, in which the user and the network storage service provider will generate the integrity metadata of the corresponding original data block respectively. All of that reach a consensus on the matter by means of the use of the blockchain technique.

Other approaches solve some problems in the recommendation system, such as scalability and cold start problem. For instance, the authors of~\cite{zhao2010user} implements a user-based collaborative filtering algorithm on a distributed cloud computing platform that is Hadoop to solve the scalability problem of the collaborative filtering method. Besides, the authors of~\cite{meng2014kasr} propose a keyword-Aware Service Recommendation method called KASR. They also present a personalized service recommendation list and keywords used to indicate user preferences. A user-based collaborative filtering algorithm is adopted to generate the recommendations. They implemented KASR on Hadoop with real-world data sets to improve its scalability and efficiency in a big data environment.
Furthermore, in~\cite{wang2020item} proposed a novel approach based on item-based $CF$ use of BERT~\cite{devlin2018bert} to help understand the items and work to show the connections between the items and solve problems that are related to the traditional recommender system as cold start. This experiment was performed with an actual data set large scale with a whole cold start scenario, and this approach has overtaken the popular Bi-LSTM model. It used the item title as content along with the item token to solve the cold start problem. The approach also further identifies the interests of the user.

Other approaches consider recommending products that are in line with the user's interests without being affected by the problems faced by the recommendation system mentioned above and the problem of data sparsity. For instance in~\cite{ferreira2020recommendation}, a product recommendation system is proposed where an autoencoder based on a collaborative filtering method is employed. The experiment result shows a very low Root Mean Squared Error (RMSE) value, considering that the users' recommendations are in line with their interests and are not affected by the data sparsity problem as the datasets are very sparse.

In e-commerce, user data and purchasing behavior play an important role~\cite{olbrich2011modeling,qiu2015predicting}.However, in our scenario we are totally agnostic about the customer behavior. The company \textit{Midiadia} does not provide complete e-commerce solutions, but gives enriched catalogues to e-commerce platforms. Consequently, \textit{Midiadia} has not information about the customers interactions, habits or any kind of profiling. To the best of our knowledge, no other study provides a solution to this problem (recommending a similar product) taking exclusively into account the product information: ingredients, size, packaging, health messages, allergens, etc. All this consideration without going back to the customer data, depends only on the product description, such as name, brand, ingredients, legal name, and size; likewise, other data helps to know that the product is also healthy, such as sugars, fats, carbohydrates and excluding all the contents that can cause allergies. Our proposition fills an exciting void for many e-commerce dominants.

\subsection{Representation Models}
\label{sec:Representation Models}
Regarding document representation models, we provide some representation models regarding the techniques used in this paper. We start with simple techniques such as Bag-Of-Words, TF-IDF.

Frist, Bag-Of-Words (a.k.a. BOW~\cite{zhang2010understanding, jiang2020understanding}) is a basic, popular, and most straightforward approach among all other feature extraction methods. It is used to create document representations in Natural Language Processing (NLP)~\cite{chowdhury2003natural} and Information Retrieval (IR)~\cite{chowdhury2010introduction}. The text is represented as a bag that contains many words. It forms a word presence feature set from all the words of an instance. The method does not care how often the word appears or the order of the words; the only thing that matters is whether the word is in the word list. It is generally used to extract features from text data in various ways. A bag of words is the presentation of text data. It specifies the frequency of words in the document. A feature generated by bag-of-words is a vector where n is the number of words in the input documents vocabulary.
Second, TF-IDF~\cite{husain2020critical} short for term frequency–inverse document frequency, is a technique that can be used as a weighting factor not only in IR solutions but also in text mining and user modeling.
This method, as in the bag-of-words model, counts how many times a word appears in a document. However, words which are repeated so many times like the stopwords (\textit{the}, \textit{of},...) are penalized with this technique because of the \textit{inverse documentary frequency} weighting. Here, the more documents a word appears in, the less relevant it is. Therefore, a word that is distinctive and frequent will be high-ranked if it appears in the query introduced by the user.

On the other hand, Word embedding is a term used for the representation of words for text analysis~\cite{lai2016generate,li2018word,hira2015review,wang2016hybrid}. It also maps of words in vectors of real numbers using the neural network, the probabilistic model, or the dimension reduction on the word co-occurrence matrix.
Word embeddings are also very useful in mitigating the curse of dimensionality, a very recurring problem in artificial intelligence~\cite{yin2018dimensionality}. Without word embedding, the unique identifiers representing the words generate scattered data, isolated points in a vast sparse representation~\cite{xing2015normalized}. With word embedding, on the other hand, the space becomes much more limited in terms of dimensionality with a widely richer amount of semantic information~\cite{ghannay2016word}. With such numerical features, it is easier for a computer to perform different mathematical operations like matrix factorization, dot product, etc. which are mandatory to use shallow and deep learning techniques.

Regarding word embedding, unfortunately, the representation of meaning with different symbols cannot orchestrate the same meaning as words. Early attempts solved this problem by clustering words based on the meaning of their endings and representing the words as high-dimensional spaced vectors. A new idea was recently proposed inspired by the neural network language model, and the model proposed is known as Word to Vector (word2vec)~\cite{mikolov2013efficient}. These embeddings are easy to work with since the vectors can be manipulated by many algorithms like dimensionality reduction, clustering, classification, similarity searching, and many more.

Two models generate the representation of word2vec have been presented in order to produce such dense word embeddings: the Continuous Bag of Word (CBOW) model~\cite{wang2017two} and the Skip-Gram model~\cite{guthrie2006closer, lazaridou2015combining}. Each of the two models train a network to predict neighboring words. Suppose that a sequence of tokens $(t_1,\dots,t_n)$is provided. The CBOW model, first randomly initializes the vector of each word and then using a single layer neural network whose outcome is the vector of the predicted word, optimizes the original guesses. One can easily understand that the size of the Neural Network controls the size of the word vector. The Skip-gram model uses the word, in order to predict the context words. After explaining the meaning of Word2Vec, however, the goal of doc2vec is to create a digital representation of the document, regardless of its length. But unlike words, documents don’t come in logical structures like words. In~\cite{le2014distributed} they used Word2Vec template and added below paragraph id to build doc2vec.

\begin{table*}[htpb] \centering
\caption{Extract of the MDD-DS}
\begin{tabular}{p{21pt}p{45pt}p{50pt}p{28pt}p{40pt}p{30pt}p{50pt}p{48pt}p{35pt}p{15pt}p{15pt}} \hline
\multicolumn{1}{c}{\boldmath{$EAN$}} &\boldmath{$Category$} &\boldmath{$Subcategory$} &\boldmath{$Variety$}&\boldmath{$Brand$} &\boldmath{$Name$}&\boldmath{$Ingredients$}&\boldmath{$Legal Name$}&\boldmath{$Servings$}&\boldmath{$Size$} &\boldmath{$Unit$}\\ \hline
{10590} & Fresh & Fish \& Shellfish   & Other  & Generic Midiadia & Congrio & Conger conger & Congrio & 1 & 330 & ml\\ \hline
{84107} & Beverages & Beers & Lager & Moritz &  Cerveza moritz & Beer. 5.4\% vol. alc.& Ron & 3 & 500 & g \\\hline
$\vdots$ & $\vdots$ & $\vdots$ & $\vdots$ & $\vdots$ & $\vdots$ & $\vdots$ & $\vdots$ & $\vdots$ & $\vdots$ & $\vdots$\\\hline
{843654}& Snacks and nuts & Nuts & Seeds & Facundo & Gian Seeds  & Sunflower seeds and salt (4\%)& Roasted and salty & 2 & 250 & g\\ \hline
\end{tabular}  
\label{tab:mdd-ds} 
\end{table*}

\section{Dataset}
\label{sec:dataset}

The data set used in this paper was provided by \textit{Midiadia}, a Spanish company which works to convert textual information in the product package into product category and product attributes by mixing automated natural language processing and manual annotation.  The \textit{Midiadia} DataSet (MDD-DS) is taxonomy where  the 3 upper levels are called Category, Subcategory and Variety. Every product in MDD-DS includes; the taxonomy position, i.e. values for Category, Subcategory and Variety as well as a set of product attributes. e.g.  name, ingredients,  legal name, brand, product size, etc., as shown the extract of real data in~\autoref{tab:mdd-ds}. We have also used these product components before in \cite{hafez2018comparative, hafez2021classification} to provide a solution to automatically categorize the constantly changing products in the market, which is the first part of our investigation.

\begin{itemize}
\item {\it 'European Article Number'} (\textit{EAN}) is an internationally recognized standard that describes the barcode and numbering system used in world trade to identify a specific product that is specifically packaged and has a specific manufacturer in retail.
\item {\it 'Category'}, {\it 'Subcategory'}, and {\it 'Variety'} are a hierarchy and can be displayed by a company as catalog organization levels in the classification. The companies manufacture the products and each company has an identifying name and is listed as the {\it brand}.
\item In addition, there are some properties compatible with the EU regulation~\cite{eu_law}, for example, name, legal name and ingredients, as indicated in~\autoref{tab:data_fields}. 
\item 'Servings' is a number that determined based on the amount of product and is sufficient for how many people.
\end{itemize}

\begin{table}[H]
    \centering
    \caption{Product attributes in the  dataset}
    \begin{tabular}{p{50pt} p{20pt} p{190pt}}
    \hline
         \textbf{Field} &\textbf{levels}& \textbf{Description} \\
         \hline
         \textit{EAN} &  & Unique Product Number \\
         \textit{Category} & 16 & 1st Level Category  \\
         \textit{Subcategory} & 62 & 2nd Level Category \\
         \textit{Variety} & 159 & 3rd Level Category, referred to as \textit{variety}\\
         \textit{Brand} & 3,015 & Product Brand \\
         \textit{Name}& 11,139 & Product Customary Name \\
         \textit{Legal Name}  & 8,442 & Official product denomination regarding the European Union provisions \\
		    \textit{Ingredients}&  & List of Ingredients in the product \\\hline
    \end{tabular}
    \label{tab:data_fields}
\end{table}

In addition, \textit{Midiadia} supported us with two versions of MDD-DS to implement recommendation systems and cover all the company’s requirements. The basic version which was called MDD-DS1, contained all the above information plus some information related to the nutrition table, such as \textit{sugar} and \textit{fat}, and some \textit{messages} on the product packaging such as the \textit{sugar-free} or the \textit{free gluten} and other messages on the cover of the product. Of course, these messages are placed according to the components of each product, as shown in~\autoref{tab:mdd-ds1}. 
\begin{table}[H] \centering
\caption{Extract of the MDD-DS1}
\begin{tabular}{cp{21pt}p{30pt}p{70pt}p{90pt}p{15pt}} \hline
\multicolumn{1}{c}{\boldmath{$EAN$}}&\boldmath{$fat$}&\boldmath{$sugar$} &\boldmath{$message$} &\boldmath{$message$} & $\dots$ \\ \hline
{10590} & 2.8 & 0 & without sugar& & $\dots$ \\ \hline
{84107} & 4.3 & 8 & & Room temperature& $\dots$  \\\hline
{$\vdots$} & $\vdots$ & $\vdots$ & $\vdots$ & $\vdots$ & $\vdots$ \\\hline
{843654} & 2.18 & 30 & & Room temperature& $\dots$ \\ \hline
\end{tabular}  
\label{tab:mdd-ds1} 
\end{table}
The extended version which was named MDD-DS2, contained all the above information besides the characteristics of the \textit{Brand Type} and \textit{Brand attributes}, and the \textit{price} was also added randomly besides more information about the nutrition table such as \textit{carbohydrates} ($Carbs$), \textit{dietary fiber} $(df)$, and \textit{a percentage of saturated fat} $(sf)$ and \textit{good fat} $(gf)$, \textit{protein} $(pn)$ and \textit{salt} $(sa)$. It also contains allergens such as \textit{soy}, \textit{fish}, \textit{eggs}, \textit{nuts}, etc., as characteristics that will be mentioned in detail and how they are used in our research, as shown in~\autoref{tab:mdd-ds2}.

\begin{itemize}
\item{\it 'Carbohydrates'} are considered one of the three main food categories and a source of energy, and they are also basically sugars and starches that the body breaks down into glucose (we can say that it is a simple sugar that the body can use to nourish its cells).
\item{\it 'Dietary Fiber'} is part of the food that has been separated from plants and cannot be completely broken down by human digestive enzymes.
\end{itemize}

\begin{table}[H] \centering
\caption{Extract of the MDD-DS2}
\begin{tabular}{p{25pt}p{66pt}p{70pt}p{19pt}p{8pt}p{8pt}p{7pt}p{7pt}p{7pt}p{7pt}p{7pt}} \hline
\boldmath{$EAN$} &\boldmath{$Brand Type$}&\boldmath{$Brand attribute$}&\boldmath{$Carbs$}&\boldmath{$df$} &\boldmath{$sf$}&\boldmath{$gf$} &\boldmath{$pn$} &\boldmath{$sa$}\\ \hline
{10590} &manufacturer& standard& 2.8 & 0& 0& 0&10&2\\ \hline
{84107} & manufacturer& without gluten& 4.3 & 8& 15& 0&4&10 \\\hline
{$\vdots$} & $\vdots$ & $\vdots$ & $\vdots$ & $\vdots$ & $\vdots$& $\vdots$ & $\vdots$ & $\vdots$ \\\hline
{843654} &White & standard& 2.18 & 30& 40& 7 &5&6\\ \hline
\end{tabular}  
\label{tab:mdd-ds2} 
\end{table}

\section{Methodology Overview}
\label{sec:Methodology}

Taking into account out dataset, the proposed recommender systems does not have information about user's history so that $CF$ should be excluded. An hybrid item-based $CF$ is designed for the specific scenario of finding similar products to a source product $P$ where similarly will be defined according to, first, the $Variety$ of the product in the MDD taxonomy and, second, other attributes of the product. The alternative product to $P$ will be a product in the same $Variety$ which moreover meets other similarity requirements over the product attributes. Three similarity approaches have been defined: (i) Product Composition (PRO-COM), where similarity is scored according to product composition (ingredient, name, legal name, etc.); (ii) Package-based (PK-BD), where similarity is scored according to the size of the product chosen by the user; and (iii) Health-based (HTH-BD), where similarity is scored according to a healthy grade by using the product nutrition table. The recommendation methodology considers allergens apart from these three similarity approaches as follows.
In MDD-DS, several product attributes are related with allergens: (\textit{Nuts}, \textit{egg}, \textit{hazelnuts}, \textit{fish}, \textit{sulfates}, \textit{peanuts}, \textit{mollusks}, \textit{lupine}, \textit{gluten}, \textit{mustard}, \textit{soy}, \textit{crustaceans}, \textit{milk and its derivatives including lactose}, \textit{sunflower seeds} and \textit{sesame}).
Allergens are considered pre-conditions for suggesting an alternative product, that is, if the user-chose a product which includes sugar, water and nuts), the allergen precondition for the alternative products is possibly containing nuts but not other allergen. So, the alternative product may nuts or not, but it should not contain other allergens.
\begin{figure}[h] \centering
\includegraphics[scale = 0.4]{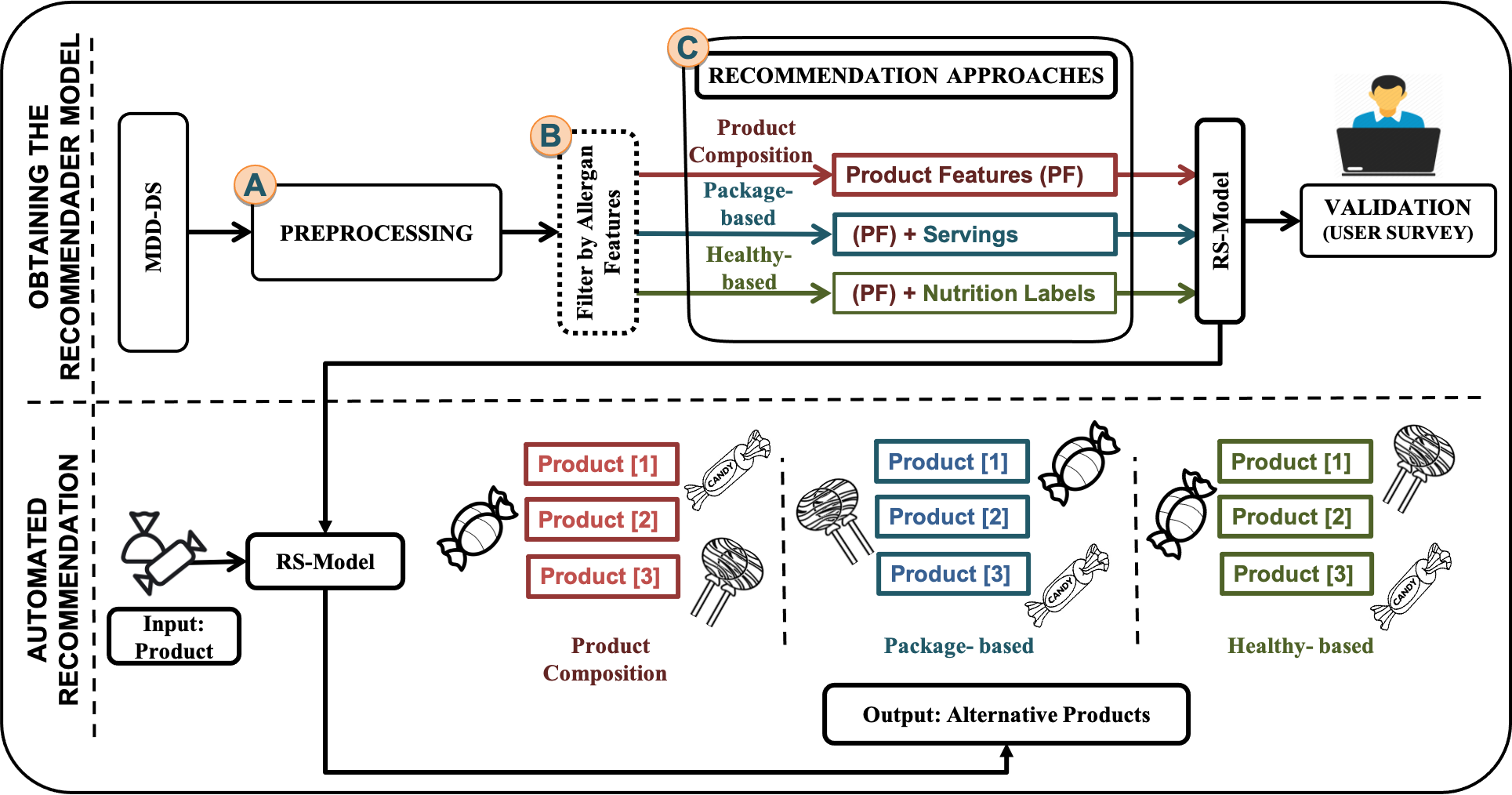} \caption{\label{fig:model0}Description of the proposed model definition and evaluation.}  
\end{figure}

The proposed methodology is shown in~\autoref{fig:model0}. In order to obtain the model a training set is defined in order to obtain the recommender model with the following steps: (A) MDD-DS is preprocessed; (B) for every product $P$ the dataset is filtered by allergen preconditions; (C) the three similarity scores are obtained  (PRO-COM, PK-BD, and HTH-BD). Then at the bottom of the model is the automated recommendation when the user selects the product. The recommendation system recommends an alternative based on the three approaches. A survey is conducted to consider the users in the three approaches.
 
\begin{figure}[h] \centering
\includegraphics[scale = 0.60]{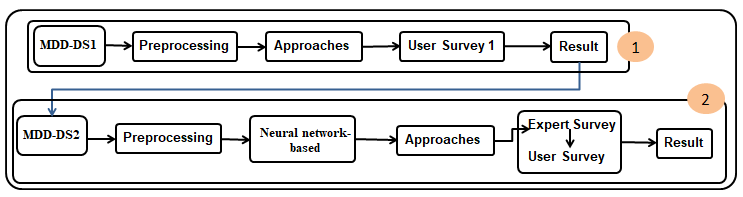} \caption{\label{fig:model}More details on description of the proposed model definition and evaluation.}  
\end{figure}

To implement the recommendation, we carried out collaborative filtering as a first model. Then we add more features and a neural network solution to improve our results. The~\autoref{fig:model}, illustrate the strategy of this paper. First model, The dataset (MDD-DS1) is analyzed by preprocessing. Three approaches were then developed, which are (PRO-COM, PK-BD, and HTH-BD) by collaborative filtering. A survey is carried out to take the users' opinions in the three approaches.

In the second model, the approaches are redeveloped based on user feedback. We added more features and more filters, like filtering by allergen features. We added a neural network solution to improve our results. Therefore, the company extends the dataset, called (MDD-DS2), to contain additional features to develop the approaches, so the data is analyzed through preprocessing. A neural network is built on the products. Then, it extracts the product as a vector and compares it to the rest of the products using similarity techniques and then makes the approaches (PRO-COM, HTH-BD, and PK-BD).

All three approaches take allergens features into account, which means that as the same explain above, if the product is, for example, nut-free, the alternative products are too. Then the approaches are sent to an expert by the company for evaluation. This has indicated that the modification is suitable for the company's requirements. Hence, a questionnaire was published for users to evaluate the recommendation system after these modifications. The last thing was a comparison of the evaluation of the users.

\section{Recommendation System based on Item-based Collaborative Filtering (RS-CF)}
\label{sec:RSnonML}

\begin{figure}[h] \centering
\includegraphics[scale = 0.40]{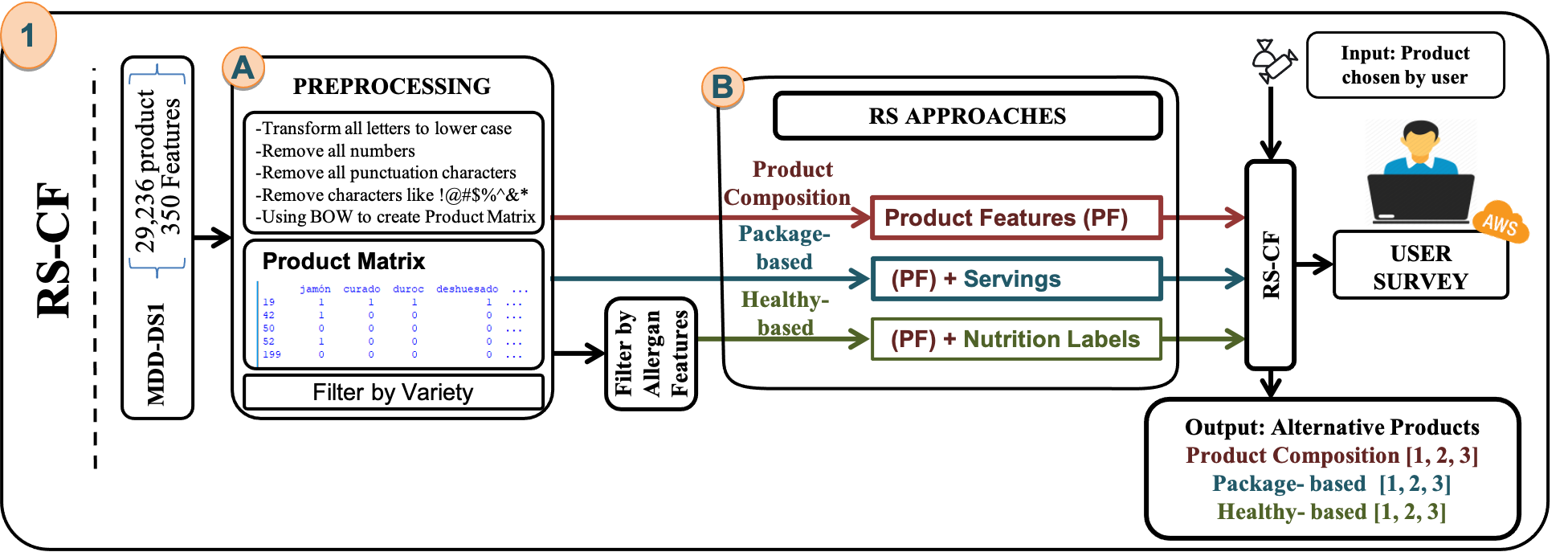} \caption{\label{fig:model1}The modeling methodology for RS-CF.}  
\end{figure}

This paper proposes a methodology to develop RS-CF for the retail sale of products. Three recommendation methods have been developed, each of which recommends alternative products to help the user obtain the product of interest. Our solution implementation takes the data (shown in the~\autoref{sec:dataset}) for each approach as the input control variables. Alternative products are then recommended for each approach and then presented to the user to choose the right product for him and evaluate RS-CF. The modeling methodology consists of 2 main steps as show in~\autoref{fig:model1}: (A) data pre-processing; (B) build the RS-CF approach; the RS-CF was done in three ways, namely: (i) Product Composition (PRO-COM) approach, where similarity is scored according to product component (ingredient, name, legal name, etc.); (ii) Package-based (PK-BD) approach, where similarity is scored based on the PRO-COM result besides the size of the product chosen by the user; and (iii) Health-Based (HTH-BD) approach, where the similarity is scored according to the PRO-COM result and taking into account that the allergen information is being considered along with a healthy degree using the product nutrition table.
In order to evaluate the RS-CF approaches by the user, we conducted a survey that includes many of the products and similar products.

It begins with PRO-COM and PK-BD approach to performing preprocessing, and building the product matrix. The algorithm made in both approaches will be explained similar with the added feature of PK-BD approach.

\subsection{Preprocessing}
\label{sec:Preprocessing}

The PRO-COM and PK-BD approaches are the first two RS-CF approaches that use the first data release (MDD-DS1). MDD-DS1 comprises 29,236 products, so the data was processed and cleaned by removing the empty rows from the variable {\it name} and {\it serving} as well. The EAN column is then scanned for duplicates and removed. Next, we ignore that the variety contains less than four products; this number is required to implement the algorithm~\footnote{The main focus of this investigation is when the primary product is not found. Therefore, keep at least three alternatives of the same variety.}. Therefore, the number of products after cleaning the data is 20,371, we mentioned the last steps in Algorithm \ref{alg:preprocess1} as a $Cleaning$(MDD-DS1) step. The data is then preprocessed by extracting all the words for the attributes \textit {name}, \textit {legal name}, and \textit {ingredients}. Consider a Corpus $C$ of each product $p$, $C$($p$[Name], $p$[Legal Name], $p$[Ingredients]).
That means that the three attributes are combined in a single text to describe the product $des(p)$. This description was obtained ($des(p)$) after cleaning the product $Clean\_p$ by following these steps:
(i) transform the parentheses into space; (ii) the numbers, stopwords, punctuation, and extra spaces are removed; (iii) all letters are converted to lowercase; and (iv) duplicate strings are removed.
Algorithm \ref{alg:preprocess1} shows all the preprocessing steps for PRO-COM and PK-BD.

\begin{algorithm}[h]

\caption{RS-CF: PRO-COM and PK-BD preprocessing pseudocode \label{alg:preprocess1}}
{\small
	\begin{algorithmic}[1]
\Procedure {Preprocess}{MDD-DS1}
\State $Cleaning$(MDD-DS1)\; 
\State  $product\_words[] \gets$ new\_list($m$) \;
\State  $all\_products\_words \gets$ new\_vector (0) \;
\For{i  $\gets$ 1 : m} 
\State $p \gets$ MDD-DS1$[i,\ ]$\;
\State $des(p)$ $\gets$ C($p$[Name], $p$[Legal Name], $p$[Ingredients])\;

\State $des(p)$  $\gets$ $Clean\_p$ ($des(p)$)\;
\State $product\_words[i]$  $\gets$ $des(p)$\;
\State $all\_products\_words$ \;
\State $\gets$ $all\_products\_words$  $\cup$ $des(p)$\;
\EndFor
\EndProcedure
\end{algorithmic}}
\end{algorithm}

Thus, the words are divided and a vector of words is created for $product\_words(p)$, an example is shown in~\autoref{tab:wordlist2}. 
\begin{table}[h] \centering
\caption{Examples of $product\_words$ for every $p$}
\begin{tabular}{ll} \hline
	\boldmath{$p$} \textbf{id} & {\textbf{Product-vector}} \\
	\hline
	1 & ['parsley', 'fresh', 'leek', ...] \\
		$\vdots$ & $\vdots$ \\
	218 & ['milk', 'skimmed', 'leek', ...]\\
	$\vdots$ & $\vdots$ \\
	$29,167$ & ['oil', 'parsley',  'lemon', ...]\\\hline
	
\end{tabular}  
\label{tab:wordlist2}
 \end{table} 
We obtain $all\_products\_words$ unique tokens/words extracted from the corpus $C$($p$[Name], $p$[Legal Name], $p$[Ingredients]), which is the different meaningful tokens in the dataset after preprocessing. Therefore, $all\_products\_words$ contains 10,707 unique tokens, an example~\footnote{We have translated $product\_words(p)$ and $all\_products\_words$ to make it readable.} shows in the~\autoref{tab:words2}. 
Let $\vec{t}$ be the $n$-dimensional vector obtained from $all\_products\_words$ such that $\vec{t} = (t_1,\dots,t_n)$ and $\forall\;k\in [1,\dots,n],\;t_k$ is a string $\in$ $all\_products\_words$ and $N=\dim(all\_products\_words)$. The $N$ tokens will form $des(p)$ and the count vector size in product matrix $X$ will be given by $MxN$.

 \begin{table}[h] \centering
\caption{Extract from $all\_products\_words$. \label{tab:words2}  }
\begin{tabular}{p{\columnwidth}} 
	\hline
['parsley', 'fresh', 'leek', 'raw', 'cauliflower', 'raw', 'thistle', 'cynara', 'cardunculu', 'panettone', 'flour', 'wheat', 'kind', 'raisin', 'egg', 'butter', 'sugar', 'orange','candied','peel','lemon','syrup', 'glucose', 'fructose', 'regulator',  'acidity', 'acid', 'citric', 'water', 'yolk', 'yeast', 'bakery', 'emulsifier', 'mono', 'diglyceride', 'fatty', 'salt', 'milk', 'skimmed', 'powder', 'flavoring', 'preservative', 'sorbic', 'plum', 'conger eel', 'conger', 'canon', 'valerianella', 'locusta', 'fruit', $\dots$]\\
\hline
\end{tabular}  
\end{table} 

The product matrix $X[m,n]$ is a $MxN$ matrix, where each row $M$ represents a product $p$ of the MDD-DS1 so that $M$ is the total number of products; and each column $N$ represents a token $t_i \in \vec{t}$. 
The next step is to use BOW to rate the words on each product. The goal is to convert each free text product into a vector that we can use as an RS model input. Since we know that the vocabulary in $all\_products\_words$ contains 10,707 words, we can use a fixed-length document-representation of 10,707, with a position in the vector to score each word. The simplest scoring method is to mark the presence of words as a boolean value, 0 for absent, 1 for present. Using the arbitrary order of $all\_products\_words$ listed above in our vocabulary $des(p)$, we can loop through the products and convert them to a binary vector, as shown in~\autoref{tab:prod_mat}.

\begin{table}[h] \centering
\caption{Example of a \textit{product matrix}.  \label{tab:prod_mat} }
{\small
\begin{tabular}{llllll}\hline 
	\boldmath{$p$} \textbf{id} &\textbf{'parsley'} &\textbf{'fresh'} & \textbf{'leek'} & $\cdots$ &\textbf{'oil'}\\ \hline
{1}& 1 & 1 & 1 & $\cdots$ & 0 \\ \hline
$\vdots$ &$\vdots$ & $\vdots$ & $\vdots$ & $\vdots$ & $\vdots$\\ \hline
{29,167} & 1 & 0 & 0 & $\dots$ & 1 \\ \hline
\end{tabular} }
\end{table} 

\subsection{RS-CF: Product Composition Approach}
\label{sec:procom}
Product composition (PRO-COM) is the main approach upon which the recommendation system is built. PRO-COM is built to obtain the alternative product based on the similarity ratio. A product matrix is used and these steps are followed to build PRO-COM approach. Let \textit{Z} a \textit{variety}. Let $\vec{d}$ be a $|Z|$-dimensional vector. Here, $\vec{d} = (d_1,\dots,d_{|Z|})$ where $d_i = d(q_i,q_j)$ denoting the following distance between the products $q_i$ and ($q_j\in Z$). 

The product $p$ is calculated by getting the absolute value of the difference considering each column of the product matrix ($all\_products\_words$) and then adding up all the distances as shown in~\autoref{eq:gen_rec1}. When a product $p\in Z$ that is not available, the RS-CF (PRO-COM) recommended $p_a\in Z$ alternative product would be obtained.
If there is more than one $p_a$ value $\vec{p}_a = (p_{a_1}, \dots, p_{a_t})$, a $t$-dimensional vector $\vec{p}_a$ is created being $t$ the number of alternatives. The alternative products given ($\vec{p}_a$) will be those that have the lower distance to the product $p$ as show that in~\autoref{eq:gen_rec_alt} and can be seen an example of a distance vector $\vec{p}_a = (p_{a_1}, \dots, p_{a_t})$ in \autoref{tab:genvector}.  

\begin{equation}
\label{eq:gen_rec1}
d(q_i,q_j) = \sum_{k = 1}^{n} |X[q_i,k] - X[q_j,k]|\;/\;q_i,q_j \in Z
\end{equation}

\begin{equation}
\label{eq:gen_rec_alt}
p_a = p_{a_i} \;\; / \;\; d(p,p_{a_i}) = \min_{\substack{\: \forall i = 1, \dots, |Z|\\ p \neq q_i}} \: \{d(p,q_i)\} \;/\;p,q_i \in Z
\end{equation}

\begin{table}[h]
\centering
\caption{Example of a distance vector $\vec{p}_a = (p_{a_1}, \dots, p_{a_t})$ for PRO-COM.}
\begin{tabular}{ccccc}
\cline{1-5}
\multicolumn{1}{c}{\boldmath{$EAN$}} & {50113009} & {76287837} &$\cdots$ & {84240702}\\
\hline
\textbf{distance} & 0 & 5 & $\cdots$ & 1,400\\
\hline
\end{tabular}
\label{tab:genvector}
\end{table}

\subsection{RS-CF: Package-based Approach}
\label{sec:pkg_app}

The PK-BD approach is to offer alternative products with product size in mind, so PK-BD approach adds more condition using an additional feature called \textit{serving} (size per person). In the same time, taking the PRO-COM distance result into account.
They are compared again to an unavailable product $p$ but with regards to the package size. 
Each product $p$ of the \textit{variety} $Z$, a distance between the product $p$ is calculated the absolute value of the package size difference. 
Then, the alternative products $p_a$ given will be those that have the lower distance with respect to the product $p$. Let $\vec{s}$ be the vector that contains the package size of the products in $Z$ ($|Z|$-dimensional vector). Hence, $\vec{s} = (s_1,\dots,s_{|Z|})$. Let $d_s(q_i,q_j)$ the following distance between the products $q_i$ and $q_j$ according to their package size as shown in~\autoref{eq:lot_rec}. 

\begin{equation}
\label{eq:lot_rec}
d_s(q_i,q_j) = |s[q_i] - s[q_j]| \;/\;q_i,q_j \in Z
\end{equation}

Considering the PK-BD approach, if a product $p\in Z$ is not available there are two steps to follow in order to get the alternative product $p_a\in Z$; (i) First of all, the PRO-COM distance is taken into account by applying \autoref{eq:gen_rec_alt}.
(ii) Next, the package size distance is additionally applied to the products in vector $\vec{p}_a|_{p} = (p_{a_1}|_{p}, \dots, p_{a_t}|_{p})$ in order to select the alternative product $p_a$ to be offered to the user as show that in~\autoref{eq:lot_rec_alt}.
If there is more than one $p_a|_l$ value, a $u$-dimensional vector $\vec{p}_a|_l = (p_{a_1}|_l, \dots, p_{a_u}|_l)$ is created being $u$ the number of alternatives.
An example of a matrix having two distance vectors taken into account the criterions selected is shown in~\autoref{tab:lotmatrix}.

    \begin{equation}
    \label{eq:lot_rec_alt}
    p_a|_{l} = {p_{a_i}}|_{l} \;\; / \;\; d_s(p,{p_{a_i}}|_{l}) = \min_{\substack{\: \forall q_{i} \in \vec{p}_a\\ p \neq q_i}} \: \{d_s(p,q_i)\} \;/\;p,q_i \in Z
    \end{equation}

\begin{table}[h]
\centering
\caption{PK-BD approach: Example of a matrix with distance vectors as columns after sorting the products.}
\begin{tabular}{ccc}
\cline{1-3}
\multicolumn{1}{c}{\boldmath{$EAN$}} & \textbf{distance (PRO-COM)} & \textbf{distance (PK-BD)}\\
\hline
{84862238} & 2 & 3\\
\hline
{84312155} & 2 & 6\\
\hline
{54487505} & 5 & 0\\
\hline
$\vdots$ & $\vdots$ & $\vdots$\\
\hline
{74410953} & 75 & 4\\
\hline
\end{tabular}
\label{tab:lotmatrix}
\end{table}

\subsection{RS-CF: Health-based Approach}
\label{sec:helthapp}
The health-based approach (HTH-BD) is the tricky one to consider recommending health products to the user based on their choices. The most common nutritional table properties \textit {fats}, \textit {sugar} are used to help recommend healthy products. The cleanliness of the data mentioned in~\autoref{sec:Preprocessing} is used in addition to replacing the \textit{serving} with the additional properties, which are \textit{fats}, \textit{sugars}, so that the rows with blank values for the name of the product, sugar and fat are eliminated, so that the sugar values in the remaining products range between 1 and 1087 grams, and the fat values in the remaining products also range between 1 and 937 grams. Additionally, 13 additional columns named {\it Messages} that provide allergen information are being considered. The number of products becomes 20,259 products and 24 features after cleaning the data.

About the {\it Messages} columns, after analyzing all the tags indicating the absence of allergens $50$ different strings are obtained in the form~\autoref{tab:masseges} and stored in a vector named {\it $withoutwords$}. Here, taking into account the law of the European Union~\footnote{BOE, Regulation (EU) n. 1169/2011 of the European Parliament and of the Council} on the labeling of food products that obliges companies to report certain allergens that may endanger the health of the customer, sensitivity will be taken into account. 
 \begin{table}[h] \centering
\caption{String values obtained from the {\it Messages} columns. \label{tab:masseges}  }
\begin{tabular}{p{\columnwidth}} 
	\hline
[without colorings, without preservatives, without additives,
without transgenics, without gluten, without artificial colors,
without trans fats, without artificial flavors, without fat,
without molluscs, without lactose, without artificial preservatives,
without sugar, without egg, without milk and its derivatives, without cholesterol, without added preservatives, without added sugar,
without salt, without palm oil, without soy, without added salt, without nuts, without peanuts, without palm , without sesame, without peanuts, without sulfites , without mustard, without saturated fat,
without alcohol, without calories, without caffeine, without sweeteners, without hydrogenated fats, 
without palm oil and fat, high protein, without added phosphates, without allergen, starch free, celery free, without artificial sweeteners, without fish, without crustaceans, without glutamate, without lupins, low fat, low in energy]\\
\hline
\end{tabular}  
\end{table} 
Most likely, a law with similar objectives was previously approved in Spain in 2004 and amended in 2008~\footnote{BOE, Royal Decree 1245/2008, of July 18, which modifies the General Regulations for Labeling, Presentation and Advertising of Food Products, Approved by Royal Decree 1334/1999 of July 31}. The 50 different strings obtained previously, 17 are the relevant ones in terms of allergies. As it can see in~\autoref{tab:masseges_allergen}. After performing the necessary analysis and clarification, the information obtained from MDD-DS1 is useful to develop the RS-CF HTH-BD algorithm.

 \begin{table}[h] \centering
\caption{Allergen features in the {\it withoutwords} vector. \label{tab:masseges_allergen}  }
\begin{tabular}{p{\columnwidth}} 
	\hline
[without allergens, without gluten, without crustaceans, without egg, without fish, without peanuts, without peanuts, without soy, without milk and its derivatives, without lactose, without nuts, without celery, without mustard, without sesame, without sulphites, without lupins, without molluscs]\\
\hline
\end{tabular}  
\end{table}

Aside from the data obtained from the \textit{Ingredients} variable, the \textit{Messages} columns associated with the respective product are also obtained for each iteration. Here, for each product, the 13 \textit{Messages} columns are handled in the following way: (1) 13 columns for the current product in the iteration are obtained, with the blank columns removed. (2) To remove additional information unrelated to the allergen, values are also removed from columns that do not begin with the string \textit{"without"}. (3) The duplicate strings obtained are removed, strings are converted to lowercase.(4) The strings are divided by a point followed by a space,  substrings preceded by a comma are removed. (5) Some incorrect parsed characters (overridden characters such as \textit{$\backslash$r} and \textit{$\backslash$n} backslashes) are removed, as well as some strings with errors and full stops are removed. The word vector is constructed with the resulting string. 

As in the PRO-COM and PK-BD approaches, the $product\_words$ list is generated with the difference that here just the \textit{Ingredients} column is considered. This is, it contains a number of elements equal to the number of different products existing in the MDD-DS1 (in the HTH-BD approach, the MDD-DS1 has 20,259 elements). The vector of words belonging to each product obtained in the text string processing is stored in each element of the list after using the steps of $Clean\_p$. The list is shown in~\autoref{tab:wordlist3}. 

\begin{table}[htb] \centering
\caption{Example of the list which contains the vectors of words belonging to each product considering the \textit{Ingredientes} column (\textit{product\_words}).}
\begin{tabular}{c|c} \hline
\boldmath{$p$} \textbf{id} & \boldmath{$des(p)$}\\\hline
{1} & ["oil", "olive", ...]\\
{2} & ["oil", "olive", ...]\\
$\vdots$ & $\vdots$ \\
{20,259} & ["lettuce", "green", ...]\\
\hline
\end{tabular}  
\label{tab:wordlist3} 
\end{table}

In addition, a list called \textit{withoutlist} which stores the vector with the healthy features obtained from the \textit{Messages} columns for each product is created. Also, the vector \textit{withoutwords} stores once the different healthy features, having 50 elements. The list and the vector are shown in~\autoref{tab:withoutlist} and \autoref{tab:withoutwords}, respectively. The entire preprocessing is shown in Algorithm~\ref{alg:prep_rechealth}. It is relevant to know that, a subset comprising 17 elements of the \textit{withoutwords} vector is considered in order to check for allergens in a product, whose data about it can be accessed by indexing the \textit{withoutlist} with the index of the product in the MDD-DS1.

\begin{table}[htb] \centering
\caption{Example of the list which contains the vectors of the features included in the \textit{Messages} columns belonging to each product (\textit{withoutlist}).}
\begin{tabular}{c|c} \hline
\boldmath{$p$} \textbf{id} & \boldmath{$Messages$}\\\hline
{1} & ["without preservatives", "low fat"]\\
{2} & ["without gluten"]\\
$\vdots$ & $\vdots$ \\
{20,259} & ["without peanuts"]\\
\hline
\end{tabular}  
\label{tab:withoutlist} 
\end{table}

\begin{table}[htb] \centering
\caption{Example of the vector which contains all the different features obtained from the \textit{Mensajes} columns (\textit{withoutwords}).}
\begin{tabular}{cccc} \hline
\textbf{1} & \textbf{2} &$\cdots$ &\textbf{50}\\
\hline
"without colorants" & "without preservatives" & $\cdots$ & "low in energy"\\
\hline
\end{tabular}  
\label{tab:withoutwords} 
\end{table}

\begin{algorithm}[H]
\caption{RS-CF: HTH-BD: Preprocessing pseudocode}
\begin{algorithmic}[1]
\Procedure {Preprocess}{MDD-DS1}
\State $Cleaning$(MDD-DS1)\; 
\State  $product\_words[] \gets$ new\_list($m$) \;
\State $withoutlist \longleftarrow$ new\_list($m$)
\State $withoutwords \longleftarrow$ new\_vector (0)
\For{i  $\gets$ 1 : m} 
\State $p \gets$ MDD-DS1$[i,\ ]$\;
\State $des(p)$ $\gets$ ($p$[Ingredients])\;\Comment{\textbf{Just the \textit{Ingredientes} column is taken into account}}
\State $des(p) \longleftarrow$ $p$

\State $des(p)$  $\gets$ $Clean\_p$ ($des(p)$)\;
\State $product\_words[i] \longleftarrow$ $des(p)$
\State $m \longleftarrow$ MDD-DS1$[i,"Messages1":"Messages13"]$\Comment{\textbf{The information contained in the 13 \textit{Messages} columns is loaded}}
\State $m \longleftarrow$ \textit{remove\_empty\_strings}($m$)
\State $m \longleftarrow$ $m$[which($m$.\textit{startsWith}("without ")]\Comment{\textbf{Just the values in the $m$ vector which start with the string "without " are loaded}}
\State $m \longleftarrow$ \textit{remove\_duplicates}($m$)
\State $m \longleftarrow$ \textit{transform\_into\_lowercase}($m$)
\State $m \longleftarrow$ \textit{split}($m$,"[.] ")\Comment{\textbf{The strings are splitted by a dot followed by a space}}
\State $m \longleftarrow$ \textit{split}($m$,"[,].*")\Comment{\textbf{The strings are splitted by a comma, removing what is after it}}
\State $m \longleftarrow$ \textit{remove\_malformed\_strings}($m$)
\State $m \longleftarrow$ \textit{remove\_full\_stops}($m$)
\State $withoutwords \longleftarrow$ $withoutwords$ $\cup$ $m$
\State $withoutlist[i] \longleftarrow$ $m$
\EndFor
\EndProcedure
\end{algorithmic}
\label{alg:prep_rechealth}
\end{algorithm}

After processing the data to be valid for building the health-based approach. Let $\vec{g}$ be the \textit{withoutwords} vector (50 elements). Then, $\vec{g} = (g_1, \dots, g_{50})$. 
Let $\vec{a}$ be the subset of the \textit{withoutwords} vector considering allergens (17 elements). Hence, $\vec{a} \subset \vec{g} \; / \; \vec{a} = (a_1, \dots, a_{17})$. 
Let $\vec{v}$ be a the $m$-dimensional  \textit{wordvectors} list. Each element contains a vector $\vec{v}_i$. Hence $\vec{v} = (\vec{v}_1, \dots, \vec{v}_{m})$. Likewise, $\vec{v_i} = (v_{i}[1], \dots, v_{i}[d_{v_{i}}])$ where $d_{v_{i}}$ is the length of the vector contained in the $i$ element of the list $\vec{v}$. Note that $\forall\;k\in[1,\dots,d_{v_{i}}],\;v_i[k]$ is a string. 
Let $\vec{v}_s$ be a $|Z|$-dimensional subset of the $Z$ elements of the $m$-dimensional $\vec{v}$ \textit{wordvectors} list. Each element contains a vector ${\vec{v}_{s_i}}$. Hence $\vec{v}_s \subset \vec{v}$ and $\vec{v}_s = ({\vec{v}_{s_1}}, \dots, {\vec{v}_{s_|Z|}})$. Likewise, ${\vec{v}_{s_i}} = ({v_{s_i}}[1], \dots, {v_{s_i}}[d_{v_{s_i}}])$ where $d_{v_{s_i}}$ is the length of the vector contained in the $i$ element of the list $\vec{v}_s$. Note that $\forall\;k\in[1,\dots,d_{v_{s_i}}],\;v_{s_i}[k]$ is a string. Let $\vec{n}_{pl}$ be a $|Z|$-dimensional vector. Here, $\vec{n}_{pl} = ({n}_{{pl}_1}, \dots, {n}_{{pl}_{|Z|}})$ where ${n}_{{pl}_{i}} = d_{v_{s_{q_i}}}$. Each element denotes the processing level of a product. 

Let $\vec{w}$ be the $m$-dimensional \textit{withoutlist} list. Each element contains a vector $\vec{w}_i$. Hence $\vec{w} = (\vec{w}_1, \dots, \vec{w}_{m})$. Likewise, $\vec{w_i} = (w_{i}[1], \dots, w_{i}[d_{w_{i}}])$ where $d_{w_{i}}$ is the length of the vector contained in the $i$ element of the list $\vec{w}$. Note that $\forall\;k\in[1,\dots,d_{w_{i}}],\;w_i[k]$ is a string. 
Let $\vec{w}_s$ be a $|Z|$-dimensional subset of the $Z$ elements of the $m$-dimensional $\vec{w}$ \textit{withoutlist} list. Each element contains a vector ${\vec{w}_{s_i}}$. Hence $\vec{w}_s \subset \vec{w}$ and $\vec{w}_s = ({\vec{w}_{s_1}}, \dots, {\vec{w}_{s_|Z|}})$. Likewise, ${\vec{w}_{s_i}} = ({w_{s_i}}[1], \dots, {w_{s_i}}[d_{w_{s_i}}])$ where $d_{w_{s_i}}$ is the length of the vector contained in the $i$ element of the list $\vec{w}_s$. Note that $\forall\;k\in[1,\dots,d_{w_{s_i}}],\;w_{s_i}[k]$ is a string. Let $\vec{n}_h$ be a $|Z|$-dimensional vector. Here, $\vec{n}_h = ({n}_{h_{1}}, \dots, {n}_{h_{|Z|}})$ where ${n}_{h_{i}} = d_{w_{s_{q_i}}}$. Each element denotes the number of healthy features of a product. 

Let $\vec{c}$ be a $|Z|$-dimensional vector. Here, $\vec{c} = (f_1 + s_1, \dots, f_{|Z|} + s_{|Z|})$ where $c_{i} = f_i + s_i$. It stores the fat and sugar features about the products. Here, $f_i$ and $s_i$ denote, respectively, the fat and sugar quantities in grams of the product $q_i$. Let $d_{a_i} = d_a(q_i,q_j)$. This denotes the following similarity measure (taking into account allergens) of the product $q_i$ with respect to the product $q_j$ as shown in~\autoref{eq:health_step1}.


\begin{equation}
        \label{eq:health_step1}
        d_a(q_i,q_j) = \Bigg\{
\begin{split}
 & 1,\; \forall h / \vec{a}[h] \in \vec{w}_s[q_i] \implies \vec{a}[h] \in \vec{w}_s[q_j] /\;q_i,q_j\;\in\;Z \\
 & 0,\; \forall h / \vec{a}[h] \in \vec{w}_s[q_i] \implies \vec{a}[h] \notin \vec{w}_s[q_j] /\;q_i,q_j\;\in\;Z
\end{split}
\end{equation}

\noindent where $ \vec{a}[h] \in \vec{w}_s[q_i]\iff\exists\;k / \vec{w}_{s_{q_j}}[k] = \vec{a}[h],\;k \in[1, \dots, d_{w_{s_{q_i}}}]$


Being the product $p\in Z$ unavailable, the alternative product $p_a\in Z$ is obtained by following the next steps: (1) The first criterion is to consider the similarity about allergens. Thus, the alternative product $p_a|_{a}$ is selected according to that measure:
    \begin{equation}
    \label{eq:health_step1_alt}
        \begin{split}
    p_a|_{a} = {p_{a_i}}|_{a} \;\;/\;\; d_a(p,{p_{a_i}}|_{a}) = 
    \max_{\substack{\: \forall i \in 1,\dots,|Z|\\ p \neq q_i}} \: \{d_a(p,q_i)\} \;/\;p,q_i \in Z
        \end{split}
    \end{equation}
    If there is more than one $p_a|_a$ value, a $u_a$-dimensional vector ${\vec{p}_a}|_a = (p_{a_1}|_a, \dots, p_{a_{u_a}}|_a)$ is created being $u_a$ the number of alternatives.
(2) Secondly, the sum of the sugar and fat quantities are considered to select the alternative product $p_a|_{c}$ among the ones in vector $\vec{p}_a|_a$:
    \begin{equation}
    \label{eq:health_step2_alt}
    p_a|_{c} = {p_{a_i}}|_{c} \;\; / \;\; \vec{c}[{p_{a_i}}|_{c}] = \min_{\substack{\: \forall q_{i} \in {\vec{p}_a}|_a\\ p \neq q_i}} \: \{\vec{c}[q_i]\} \;/\;p,q_i \in Z
    \end{equation}
    If there is more than one $p_a|_c$ value, a $u_c$-dimensional vector ${\vec{p}_a}|_c = (p_{a_1}|_c, \dots, p_{a_{u_c}}|_c)$ is created being $u_c$ the number of alternatives.
(3) The next criterion to get the alternative product $p_a|_{h}$ (among the ones in vector ${\vec{p}_a}|_c$) is the number of healthy features:
    \begin{equation}
    \label{eq:health_step3_alt}
    p_a|_{h} = {p_{a_i}}|_{h} \;\; / \;\; \vec{n}_h[{p_{a_i}}|_{h}] = \max_{\substack{\: \forall q_{i} \in {\vec{p}_a}|_c\\ p \neq q_i}} \: \{\vec{n}_h[q_i]\} \;/\;p,q_i \in Z
    \end{equation}
    If there is more than one $p_a|_h$ value, a $u_h$-dimensional vector ${\vec{p}_a}|_h = (p_{a_1}|_h, \dots, p_{a_{u_h}}|_h)$ is created being $u_h$ the number of alternatives.
(4) The last step to get the alternative product $p_a|_{n_{pl}}$ involves the level of processing of the products selecting from the vector ${\vec{p}_a}|_h$:
    \begin{equation}
    \label{eq:health_step4_alt}
        \begin{split}
     p_a|_{n_{pl}} = {p_{a_i}}|_{n_{pl}} \;\; / \;\; \vec{n}_p[{p_{a_i}}|_{n_{pl}}] = \min_{\substack{\: \forall q_{i} \in {\vec{p}_a}|_h\\ p \neq q_i}} \: \{{\vec{n}_{pl}}[q_i]\} \;/\;p,q_i \in Z
     \end{split}
    \end{equation}
    If there is more than one $p_a|_{n_{pl}}$ value, a $u_{n_{pl}}$-dimensional vector ${\vec{p}_a}|_{n_{pl}} = (p_{a_1}|_{n_{pl}}, \dots, p_{a_{u_{n_{pl}}}}|_{n_{pl}})$ is created being $u_{n_{pl}}$ the number of alternatives:

In conclusion, the algorithm~\ref{alg:health} compares first each $q_i$ product in the variety $Z$ to the product $p$ with regards to the similar features about allergens. That similar products are then ranked considering this features in order: the sum of the fat and sugar amounts (in increasing order), the number of healthy features (in decreasing order) and the processing level (in increasing order). An example of a matrix with vectors defining each of the criterions as columns is shown in~\autoref{tab:healthmatrix}.

\begin{table}[h]
\centering
\caption{HTH-BD: Example of a matrix with the considered criterions as columns after sorting the products.}
\begin{tabular}{p{30pt}p{50pt}p{50pt}p{50pt}p{50pt}} \hline
\cline{1-5}
\multicolumn{1}{c}{\boldmath{$EAN$}} & \textbf{Similarity} & \textbf{Fat + Sugar} & \textbf{Healthy features} & \textbf{Processing Level}\\
\hline
{6431649} & 1 & 4 & 6 & 5\\
\hline
{7358802} & 1 & 4 & 3 & 0\\
\hline
{652108} & 0 & 0.47 & 0 & 1\\
\hline
$\vdots$ & $\vdots$ & $\vdots$ & $\vdots$ & $\vdots$\\
\hline
4452030 & 0 & 10.26 & 4 & 15\\
\hline
\end{tabular}
\label{tab:healthmatrix}
\end{table}

After building the three approaches, a user survey was conducted. Products and alternatives were presented according to each approach. Subsequently, the analyses of the results were compiled. We developed the approach to improve the results and meet the company's requirements. 

\begin{algorithm}[H]
\caption{HTH-BD approach of RS-CF: Algorithm pseudocode}\label{alg:health-based-approach}
\begin{algorithmic}[1]
\Procedure{Algorithm}{$p$, $\vec{a}$, $\vec{v}_s$, $\vec{w}_s$, $\vec{c}$}\Comment{$p$ is the index of the unavailable product}
\For{$i \leftarrow 1:|Z|$}
\If{$p == i$}
    \State \textrm{\textbf{continue}}
\EndIf
\State $a\_indexes \gets$ which($a$ $\in$ $\vec{w_s[p]}$)
\If{$\forall$ $a\_indexes$, $a[a\_indexes] \in \vec{w_s[i]}$}
    \State $d_a[i] = 1$
\Else
    \State $d_a[i] = 0$
\EndIf
\State ${n_h}[i] = length(\vec{w_s[i]})$
\State ${n_{pl}}[i] = length(\vec{v_s[i]})$
\EndFor
\State ${p_a} \gets$ sort(-$\vec{d}_a$, $\vec{c}$, -$\vec{n}_h$, $\vec{n}_{pl}$)\Comment{The products are sorted. The minus sign means the order is decreasing.}
\EndProcedure
\end{algorithmic}
\label{alg:health}
\end{algorithm}

\section{Recommendation System based on Neural Network-based (RS-NN)}
\label{RS-ml}

The idea of improving RS-CF is based on improving the result and considering more conditions and filtering: (1) Adding allergens' properties as a pre-condition in the recommendation for three approaches (PRO-COM, PK-BD, and HTH-BD). For example, the product includes (flour, eggs, water, nuts, and salt), so the alternative product will include free allergens, or the maximum allergens are eggs and nuts. 
(2) We also consider more conditions for three approaches based on using more additional features such as \textit{brand type}, \textit{brand attribute} and \textit{price}. 
(3) Besides considering the more characteristics of the nutritional table, such as \textit{carbohydrates}, \textit{dietary fiber}, \textit{a percentage of saturated fat}, \textit{good fat}, \textit{protein} and \textit{salt} to improve the HTH-BD approach.
(4) Rearrange the approaches of PRO-COM, then HTH-BD, then PK-BD. Also, to improve the result, we thought about using a deep neural network like Doc2vec to represent the data set and build a model to help obtain alternative products. That is why we call this model a Recommendation system based on neural networks (RS-NN). 

After that, we use many of the similarity techniques like Cosine, Jaccard, Euclidean, and Manhattan to obtain and sort similar products. Subsequently, we conduct a comparative study to determine which technique is best to sort similar products based on the experts' results.

\begin{figure}[h] \centering
\includegraphics[scale = 0.40]{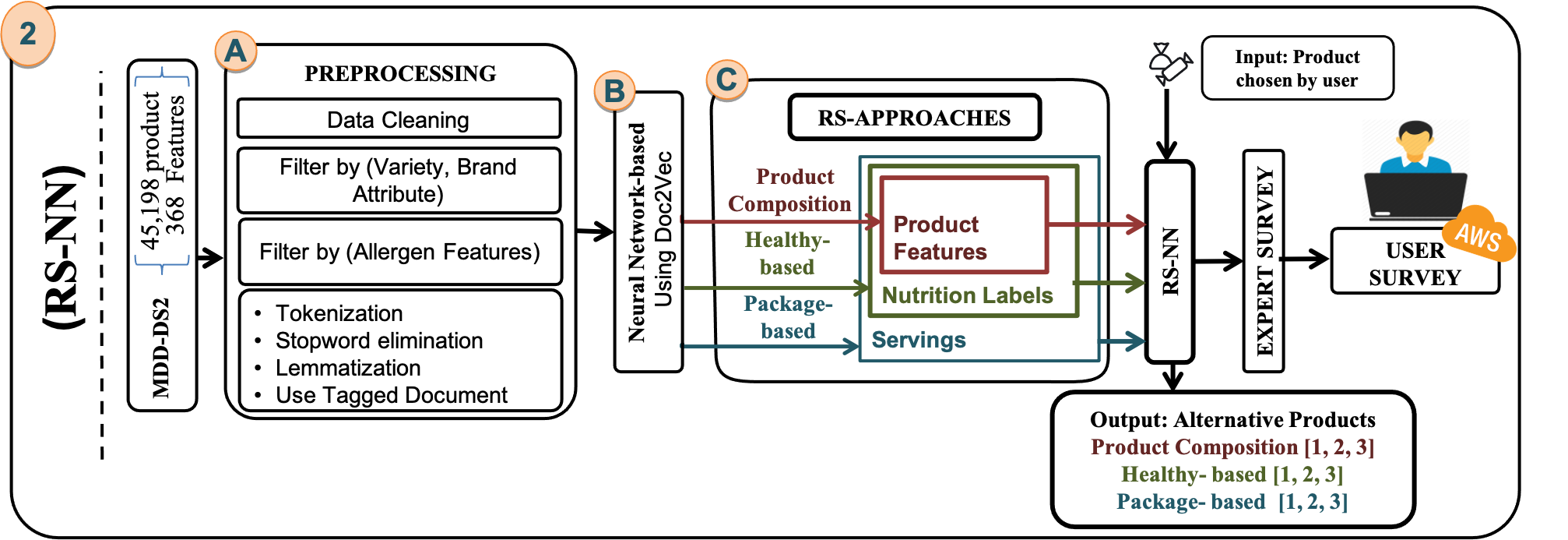} \caption{\label{fig:model2}The modeling methodology for RS-NN.}  
\end{figure}

\autoref{fig:model2} illustrates that the new model comprises three main steps: 
(A) Preprocessing the dataset using text mining,filtering,and representing the adaptive dataset with a neural network model.
(B) Using neural networks to create a model based on Doc2vec. 
(C) We apply the three RS-NN approach (PRO-COM, HTH-BD, and PK-BD). 
An expert first carries out the evaluation; then we did a user survey based on the result of the expert.

\subsection{Preprocessing}
\label{sec:pre_ml}

In order to build the RS-NN approaches, we will use some more features , such as the \textit{brand type} and the \textit{brand attributes}, as well as the addition of 16 characteristics that cause allergies (\textit{Nuts}, \textit{egg}, \textit{hazelnuts}, \textit{fish}, \textit{sulfates}, \textit{peanuts}, \textit{mollusks}, \textit{lupine}, \textit{gluten}, \textit{mustard}, \textit{soy}, \textit{crustaceans}, \textit{milk and its derivatives including lactose}, \textit{sunflower seeds} and \textit{sesame}). So, the \textit{Midiadia} extend new version of data set (MDD-DS2) which is the number of products 45,198 product and the number of features is 368 features.  To improve the approaches, the data cleansing, preprocessing, and approach building phase was reused, so blank rows were removed from each \textit{name}, \textit{brand},\textit{ brand type}, \textit{brand attributes}, \textit{variety}, \textit{ingredients}, and \textit{legal name}. Also, the elimination of the duplicate rows that have the same \textit{name} and \textit{brand} and finally the empty and duplicate rows were eliminated from the \textit{EAN} variable. The main idea of the recommendation system is to have alternatives for the product of the same \textit{variety}. Therefore, we will remove all \textit{variety} with less than the first quarter, which equates to 15 products, \autoref{tab:pr-variety} shows the products per variety after eliminate the \textit{variety} for PRO-COM approach~(\autoref{sec:firstapp_ml}).

\begin{table}[htb] \centering
\caption{Products per variety after eliminate first quarter.}
\begin{tabular}{cccccc} \hline
\textbf{Approaches}&\textbf{Count}&\textbf{Min} &\textbf{Median} &\textbf{Mean} &\textbf{Max.}\\ \hline
PRO-COM & 342 & 15 & 41 &63.7 & 501  \\
HTH-BD & 303 & 4 & 14 &26.5 & 233  \\
PK-BD & 292 & 4 & 13 &24.3 & 203  \\\hline
\end{tabular}  
\label{tab:pr-variety} 
\end{table}

The variety $Z$, $p\in Z$ that is not available, the recommended $p_a\in Z$ alternative product. After that, a product $p$ is in a specific variety called "$other$" ($p \in Z = "other"$), it is listed at a subcategory. let \textit{SC} a \textit{subcategory}, where $Z\in SC$, so the other subcategories $SC$ will be remove. Besides, let $Ba$ a \textit{Brands attributes} have a product $p \in Ba$ that is not available, the recommended $p_a \in Ba$ alternative product. Finally, it filters the products $p$ according to the allergens feature, let $af$ be allergen features. Eliminating all the products that contain more or different allergens feature $af$.

The products $p$ and $p_a$ alternative product will be preprocessed by extracting all the words for each of the following attributes: \textit{Name}, \textit{brand}, \textit{Ingredients}, \textit{legal Name}, and \textit{allergens feature} for each product $p$, $p_a$ alternative product. Consider a Corpus $C$ of ($Name$, $brand$, $Ingredients$, $legal Name$, $allergens feature$) to describe the product $des (p)$.
This description ($des (p)$) was obtained after the 3-step purge process: (1) Use the tokenize function to make a list to convert everything to lower words and separate each word $product\_words(p)$. (2) Use stopwords in Spanish to filter by stopwords such as remove [and, or, etc.], use the number filter to remove all numbers from the list. (3) Using a lemmatization step takes the tokens and divides each one into a lemma~\footnote{A lemma is basically the basic form of the word, it cuts the conjugation and declension method. For example, the word “different” would become “differ”, “running” would become “run” and “trucks” would become “truck”. Lemmatization can even change from “was” to “be” because the lemmatizer (nltk) improves vocabulary rather than relying solely on the algorithm}, similar words are then removed, 
as show in~\autoref{tab:prod_ing} step (3). 

After pre-processing, we used Tagged\_Document function~\footnote{Represents a product along with a tag, input product format for Doc2Vec, a single product, made up of words (a list of unicode string tokens) and tags (a list of tokens). Tags may be one or more unicode string tokens, but typical practice (which will also be the most memory-efficient) is for the tags list to include a unique integer id as the only tag.}  for training corpus $C$($Name$, $brand$, $Ingredients$, $legal Name$, $allergens feature$). Let $\vec{p}$ be the $product\_words[]$ list such that, each element contains a vector $\vec{p}_i$ hence $\vec{p} = (\vec{p}_1, \dots, \vec{p}_{m})$. Likewise, $\vec{p_i} = (p_{i}[1], \dots, p_{i}[l])$ where $l=\dim(product\_words[i])$, length of the vector in the $i$ element of the list $\vec{p}$. Note that $\forall\;k\in [1,\dots,l],\;p_i[k]$ is a string. Then beside each $p$, the Tagged\_Document function define the tag (the product id), which means simply the zero-based line number as shown in~\autoref{tab:taggeddocument}. 

\begin{table}[h] \centering
\caption{Examples of $product\_words$ for every $p$}
\begin{tabular}{p{60pt}p{300pt}} \hline
	\textbf{Steps} & \boldmath{$des (p)$} \\
	\hline
{Original ($p$)} & ['Wine', 'White', 'Sauvignon', 'Blanc', 'Type', 'Grape', 'Sauvignon', 'Blanc', '13', 'Vol', 'Alc'] \\
{(1)} & ['wine', 'white', 'sauvignon', 'blanc', 'type', 'grape', 'sauvignon', 'blanc', '13', 'vol', ' alc '] \\
{(2)} & ['wine', 'white', 'sauvignon', 'blanc', 'type', 'grape', 'sauvignon', 'blanc', 'vol', 'alc'] \\
{(3)} & ['wine', 'white', 'sauvignon', 'blanc', 'type', 'grape', 'vol', 'alc']\\\hline
\end{tabular}  
\label{tab:prod_ing}
 \end{table}

\begin{table}[h] \centering
\caption{The first two $des (p)$ by Tagged\_Document. \label{tab:taggeddocument}}
\begin{tabular}{p{\columnwidth}} 
	\hline
[TaggedDocument(words=['dessert', 'dairy', 'apricot', 'milk', 'whole', 'water', 'sugar', 'starch', 'modified', 'corn', 'puree', 'rice', 'stabilizer','pectin', 'oil', 'rapeseed', 'caseinate', 'sodium calcium', 'contains', 'aroma', 'natural', 'corrector', 'acidity', 'acid', 'lactic', 'food', 'infant', 'sugary', 'derivative', 'included', 'lactose'], tags = ['0']), \\
TaggedDocument (words = ['dessert', 'dairy', 'cocoa', 'milk', 'whole', 'water', 'sugar', 'thickener', 'starch', 'modified', 'corn', 'gum', 'locust bean', 'powder', 'oil', 'rape', 'caseinate', 'soda-calcium', 'contains',' corrector', 'acidity', 'hydroxide', 'sodium', 'food', 'infant', 'sugary', 'derivative', 'included', 'lactose'], tags = ['1'])]\\
\hline
\end{tabular}  
\end{table} 

\subsection{Product Representation}
\label{sec:neural}

As explained earlier the doc2vec in~\autoref{sec:Representation Models}, it shows the simplest way to convert a token to a fixed-size digital vector, as it proposed a neural network-based word representation method called Word2Vec. Give a sequence of training tokens $[t_1, t_2,\dots,t_{N-1}, t_N]$; the goal of Word2Vec is to maximize the average log probability~\cite{karvelis2018topic}.

\begin{equation}
\label{eq:wvec}
\frac{1}{N}\sum_{n = s}^{N-s}\log p(t_n|t_{n-s},\dots,t_{n+s}),
\end{equation}
where $s$ is the size of the window to preserve contextual information, the token $t_n$ can be easily be predicted using a multilabel classifier like SoftMax function:

\begin{equation}
\label{eq:softmax}
p(t_n|t_{n-s},\dots,t_{n+s})=\frac{e^{j_{t_{n}}}}{\sum_{i}{e^{j_{t_{i}}}}},
\end{equation}
where each $j_{t_{i}}$ is the ($t_i$) output value of a feed-forward neural network calculated with

\begin{equation}
\label{eq:word2vec}
j=x+hf(t_{n-s},\dots,t_{n+s};R),
\end{equation}
where $x$, $h$, $f$, and $R$ are terms for the bias between the hidden and output layers, the weight matrix between the hidden and output layers, the mean or sequence of product tokens, and the word embedding matrix, respectively. Doc2Vec extends from Word2Vec that tries to define, in this case, a continuous vector fit to a product to preserve the semantic relationship between the different products~\cite{kim2019multi}. Like Word2Vec, each token is represented by a d-dimensional continuous vector ($d <<|v|$, which is the size of the vocabulary in the $des(p)$). Furthermore, the same product $p$ is also represented by a continuous vector in the same space as the word vectors. In Doc2Vec, each product $p$ is assigned to a unique vector that is represented by a column in matrix $D$, while each token in the $des(p)$ is assigned to a unique vector that is represented by a column in matrix $T$. Therefore, the only change in the network formulation is to add $D$ to the~\autoref{eq:word2vec} as follows:

\begin{equation}
\label{eq:gen_rec}
j=x+hf(t_{n-s},\dots,t_{n+s};R,D)
\end{equation}
When the network is adequately trained, it can obtain a distributed representation of each of the products $p$. Therefore, the products were trained using three elements of the Doc2Vec model, vector size with 50 dimensions $\vec{a} = (a_1, \dots, a_{50})$, and iteration over the training set 40 times. Set the minimum word count to two to ignore words with very few frequencies.

Finally, we have a product matrix $X[m,n]$ is a $MxN$ matrix where each column $N$ represents a vector for each product $\vec{a} = ([a_1, \dots, a_{50}],tag)$, $\forall\;k\in [1,\dots,50]$, where $tag \in p id$, and each row $M$ represents a product $p$ of the MDD-DS2 so that $M$ is the total number of products $\vec{p} = (p_1, \dots, p_{m})$, as shown in~\autoref{tab:v1}.

\begin{table}[h] \centering
\caption{The vector for first product $\vec{a}(p_1)$. \label{tab:v1}}
\begin{tabular}{p{\columnwidth}} 
	\hline
array([ 0.00742837, -0.00540146, -0.14862166, -0.00862698,  0.31875622,
        0.115518,  0.00795528, -0.06915003,  0.03247217, -0.12760445,
       -0.20222402, -0.09181757, -0.02992765, -0.09429716,  0.04839283,
       -0.08727524, -0.08463322, -0.09556159, -0.01945411, -0.0644968,
        0.11707045, -0.09715877, -0.24429108, -0.08826657, -0.12004123,
       -0.17009708,  0.17322347, -0.04258763,  0.03453251, -0.19297938,
       -0.2081344,  0.23702264,  0.08457132,  0.0120729,  0.03960438,
       -0.21322013,  0.09752178, -0.03770451, -0.06469689,  0.02615795,
        0.20623626, -0.09590556, -0.00720048, -0.12926176, -0.21335329,
       -0.11945274,  0.06031954,  0.0124997,  0.27832198, -0.10382865],
      dtype=float32)\\
\hline
\end{tabular}  
\end{table} 

\subsection{Similarity}
\label{sec:simi_techq}
The RS-NN approaches used the similarity techniques such as (\textit{Cosine}, \textit{Jaccard}, \textit{Euclidean}, and \textit{Manhattan}) to calculate the distance between the product $q_i$ and $q_j$. Let $d_i = d(q_i,q_j)$, this denotes the following similarity measure taking into account \textit{variety} $Z$, \textit{brand attribute} $Ba$, and \textit{allergens features} $af$ of the product $q_i$ with respect to the product $q_j$ as show that in~\autoref{eq:similarity}.

\begin{equation} 	\begin{split}
&Cos[d(q_i,q_j)] = \frac {q_i \cdot q_j}{||q_i|| \cdot ||q_j||}\;/\;q_i,q_j \in Z, Ba \\
&Jac[d(q_i,q_j)] =\frac{|q_i\cap q_j|}{|q_i\cup q_j|}\;/\;q_i,q_j \in Z, Ba \\
&E[d(q_i,q_j)] =\sqrt{\sum_{k = 1}^{n} |X[q_i,k] - X[q_j,k]|\;^2}\;/\;q_i,q_j \in Z, Ba  \\
&M[d(q_i,q_j)] = \sum_{k = 1}^{n} |X[q_i,k] - X[q_j,k]|\;/\;q_i,q_j \in Z, Ba 
\end{split}
\label{eq:similarity}\end{equation} 

Having a product $p\in Z$ and $p \in Ba$ that is not available, the recommended $p_a\in Z$ and $p_a \in Ba$ alternative product would be obtained as follows taking into account the allergen featuresas shown in~\autoref{eq:basic_rec}, the first equation is the output from \textit{Cosine}, \textit{Jaccard}, the second one for \textit{Euclidean}, and \textit{Manhattan}. If there is more than one $CJ[p_a]$, $EM[p_a]$ value $CJ[\vec{p}_a] = (p_{a_1}, \dots, p_{a_c})$, $EM[\vec{p}_a] = (p_{a_1}, \dots, p_{a_c})$ a $c$-dimensional vector $\vec{p}_a$ is created being $c$ the number of alternatives.

\begin{equation}
\label{eq:basic_rec}
\begin{split}
CJ[p_a] = p_{a_i} \;\; / \;\; d(p,p_{a_i}) = \max_{\substack{\: \forall i = 1, \dots, |Z, Ba|\\ p \neq q_i}} \: \{d(p,q_i)\} \;\\
EM[p_a] = p_{a_i} \;\; / \;\; d(p,p_{a_i}) = \min_{\substack{\: \forall i = 1, \dots, |Z, Ba|\\ p \neq q_i}} \: \{d(p,q_i)\} \;
\end{split}
\end{equation}

\subsection{RS-NN: Product Composition Approach}
\label{sec:firstapp_ml}
The product composition (PRO-COM), where similarity is scored according to product matrix to offer alternative products. In addition, the alternatives taking into account the distance based on $d(q_i,q_j)$, they are compared to the unavailable product but with regards to the \textit{brand}, \textit{brand type}, and \textit{price}. For each product of the \textit{variety} $Z$, and $p \in Ba$ \textit{brand attribute}, a distance between the product $p$ is calculated using similarity techniques~\autoref{eq:similarity}.

Considering the PRO-COM approach, if a product $p\in Z$, and $p \in Ba$ is not available in order to get the alternative product $p_a\in Z$, and $p_a \in Ba$.
Let $\vec{b}$ be the vector that contains the \textit{brand} of the products in $Z$ ($|Z|$-dimensional vector). Hence, $\vec{b} = (b_1,\dots,b_{|Z|})$.
Beside, let $\vec{bt}$ be the vector that contains the \textit{brand type} of the products in $Z$ ($|Z|$-dimensional vector). Hence, $\vec{bt} = (bt_1,\dots,bt_{|Z|})$.
In addition, let $\vec{PR}$ be the vector that contains the \textit{price} of the products in $Z$ ($|Z|$-dimensional vector). Hence, $\vec{PR} = (PR_1,\dots,PR_{|Z|})$.

Considering of verifying the brand ${b}$ and brand type ${bt}$ in the product $p$ and that $p_a$ alternative product contains the same value for the two variables $(b, bt)$ , so we found three possibilities:
(1) The alternative product $q_j$ has the same attributes value for $(b, bt)$ of the product $q_i$. 

$\scriptstyle Pos(1) = \;$ $\forall$ ${\scriptstyle m}$ / ${\scriptstyle \vec{b}[m]}$ ${\scriptstyle =}$ ${\scriptstyle \vec{b}[q_i]}$ $ \scriptstyle \land$ ${ \scriptstyle \vec{bt}[m]}$ ${\scriptstyle =}$ ${\scriptstyle \vec{bt}[q_i]}$ ${\scriptstyle \implies}$ ${\scriptstyle \vec{b}[m]}$ ${\scriptstyle =}$ ${\scriptstyle \vec{b}[q_j]}$ $ \scriptstyle \land$ ${ \scriptstyle \vec{bt}[m]}$ ${\scriptstyle =}$ ${\scriptstyle \vec{bt}[q_j]}$ ${\scriptstyle /\;q_i,q_j\;\in\;Z, Ba}$

(2) The alternative product $q_j$ has the attribute value of one of $(b, bt)$ of the product $q_i$.

$\scriptstyle Pos(2) = \;$ $\forall$ ${\scriptstyle m}$ / ${\scriptstyle \vec{b}[m]}$ ${\scriptstyle =}$ ${\scriptstyle \vec{b}[q_i]}$ $ \scriptstyle \land$ ${ \scriptstyle \vec{bt}[m]}$ ${\scriptstyle =}$ ${\scriptstyle \vec{bt}[q_i]}$ ${\scriptstyle \implies}$ ${\scriptstyle \vec{b}[m]}$ ${\scriptstyle =}$ ${\scriptstyle \vec{b}[q_j]}$ $ \scriptstyle \lor$ ${ \scriptstyle \vec{bt}[m]}$ ${\scriptstyle =}$ ${\scriptstyle \vec{bt}[q_j]}$ ${\scriptstyle /\;q_i,q_j\;\in\;Z, Ba}$

(3) The alternative product $q_j$ does not have the same value for $(b, bt)$ of the product $q_i$.

$\scriptstyle Pos(3) = \;$ $\forall$ ${\scriptstyle m}$ / ${\scriptstyle \vec{b}[m]}$ ${\scriptstyle =}$ ${\scriptstyle \vec{b}[q_i]}$ $ \scriptstyle \land$ ${ \scriptstyle \vec{bt}[m]}$ ${\scriptstyle =}$ ${\scriptstyle \vec{bt}[q_i]}$ ${\scriptstyle \implies}$ ${\scriptstyle \vec{b}[m]}$ ${\scriptstyle \neq}$ ${\scriptstyle \vec{b}[q_j]}$ $ \scriptstyle \land$ ${ \scriptstyle \vec{bt}[m]}$ ${\scriptstyle \neq}$ ${\scriptstyle \vec{bt}[q_j]}$ ${\scriptstyle /\;q_i,q_j\;\in\;Z, Ba}$

And also to check the price, there are two options in each possibility of variables $(b, bt)$; the \textit{price} $PR$ of the alternative product $q_j$ is higher than the product $q_i$ or vice versa. Let $CJ[d_{r_i}] = CJ[d_{r}(q_i,q_j)]$ for \textit{cosine} and \textit{jaccard}, let $EM[d_{r_i}] = EM[d_{r}(q_i,q_j)]$ for \textit{euclidean} and \textit{manhattan}. This denotes the following similarity measure of the product $q_i$ with respect to the product $q_j$ as show that in~\autoref{eq:price_step1}.

\begin{equation}
        \label{eq:price_step1}
        CJ[d_{r}(q_i,q_j)] = \Bigg\{
        \begin{tabular}{l}
        ${\scriptstyle \textrm{if} \;}$ ${\scriptstyle PR[q_j]}$ ${\scriptstyle >}$ ${\scriptstyle PR[q_i] ,}$\\ ${\scriptstyle d(q_i,q_j)\;\times\; (PR[q_i]\; /\; PR[q_j])}$ \\ 
         ${\scriptstyle \textrm{if} \;}$ ${\scriptstyle PR[q_j]}$ ${\scriptstyle \leq}$ ${\scriptstyle PR[q_i],}$\\
        ${\scriptstyle d(q_i,q_j)\;\times\; (PR[q_j]\; /\; PR[q_i])}$\\
        \end{tabular}\\
        EM[d_{r}(q_i,q_j)] = \Bigg\{
        \begin{tabular}{l}
        ${\scriptstyle \textrm{if} \;}$ ${\scriptstyle PR[q_j]}$ ${\scriptstyle >}$ ${\scriptstyle PR[q_i],}$\\${\scriptstyle (PR[q_i]\; /\; PR[q_j])\; / \;\scriptstyle d(q_i,q_j)\;}$ \\
         ${\scriptstyle \textrm{if} \;}$ ${\scriptstyle PR[q_j]}$ ${\scriptstyle \leq}$ ${\scriptstyle PR[q_i],}$\\${\scriptstyle (PR[q_j]\; /\; PR[q_i]) \; / \;\scriptstyle d(q_i,q_j)\;}$\\
        \end{tabular}
\end{equation}

After check the possibilities for $p_a$ alternatives product of variables \textit{brand} and \textit{brand type} $(b,bt)$, and calculate the distance $CJ[d_{r}(q_i,q_j)]$, $EM[d_{r}(q_i,q_j)]$. Let $CJ[d_{m_i}] = CJ[d_{m}(q_i,q_j)]$ for \textit{cosine} and \textit{jaccard}, let $EM[d_{m_i}] = EM[d_{m}(q_i,q_j)]$ for \textit{euclidean} and \textit{manhattan}. This denotes the following similarity measure of the product $q_i$ with respect to the product $q_j$ as shown in ~\autoref{eq:price_step2}. 
Lastly, we will multiply the distance $CJ[d_{r}(q_i,q_j)]$, $EM[d_{r}(q_i,q_j)]$ with weight like (100, 10, 1) to help the $p_a$ alternative product's ordering.

\begin{equation}
        \label{eq:price_step2}
        CJ[d_{m}(q_i,q_j)] = \Bigg\{
        \begin{tabular}{l}
        ${\scriptstyle d_r(q_i,q_j)\;\times\; 100 ,}$ ${\scriptstyle \textrm{if} \;}$ ${\scriptstyle Pos(1)}$\\
        ${\scriptstyle d_r(q_i,q_j)\;\times\; 10 ,}$ ${\scriptstyle \textrm{if} \;}$ ${\scriptstyle Pos(2)}$\\
        ${\scriptstyle d_r(q_i,q_j)\; ,}$ ${\scriptstyle \textrm{if} \;}$ ${\scriptstyle Pos(3)}$\\
        \end{tabular}\\
        EM[d_{m}(q_i,q_j)] = \Bigg\{
        \begin{tabular}{l}
        ${\scriptstyle d_r(q_i,q_j)\; ,}$ ${\scriptstyle \textrm{if} \;}$ ${\scriptstyle Pos(1)}$\\
        ${\scriptstyle d_r(q_i,q_j)\;\times\; 10 ,}$ ${\scriptstyle \textrm{if} \;}$ ${\scriptstyle Pos(2)}$\\
        ${\scriptstyle d_r(q_i,q_j)\;\times\; 100 ,}$ ${\scriptstyle \textrm{if} \;}$ ${\scriptstyle Pos(3)}$\\
        \end{tabular}
\end{equation}

The distance is additionally applied to the products 
in order to select the alternative product to be offered to the user as shown in~\autoref{eq:price_3}. 
If there is the output from the similarity techniques (\textit{Cosine}, \textit{Jaccard}, \textit{Euclidean}, and \textit{Manhattan})  more than one alternatives product $p_a|_b$ value, a $y$-dimensional vector $\vec{p}_a|_b = (p_{a_1}|_b, \dots, p_{a_y}|_b)$ is created being $y$ the number of alternatives.
    \begin{equation}
    \label{eq:price_3}
    p_a|_{b} = {p_{a_i}}|_{b} \;\; / \;\; d_m(p,{p_{a_i}}|_{b}) = \max_{\substack{\: \forall q_{i} \in \vec{p}_a\\ p \neq q_i}} \: \{d_m(p,q_i)\} \;
    \end{equation}

Finally, after its development, PRO-COM works on three main characteristics which are the \textit{brand}, \textit{brand type} and the \textit{price}. After obtaining a vector $\vec{p}_a|_b$, the alternative products are ordered from closest to furthest. 

\subsection{RS-NN: Healthy-based Approach}
\label{sec:secondapp_ml}
The health-based (HTH-BD) approach depends on the result of PRO-COM approach and make an equation for nutrition table features.
The HTH-BD was based on the most health-based characteristics found in the nutrition table, namely \textit{fats} $(f)$, containing a \textit{percentage of saturated fat} $(sf)$ and a \textit{percentage of good fats} $(gf)$; $(sf, gf \in f)$, \textit{carbohydrates} $(Carbs)$, and containing \textit{dietary fibers} $(df)$ and \textit{sugars} $(s)$ $(df, s \in Carbs)$, and finally \textit{salt} $(sa)$ and \textit{protein} $(pn)$. 

Eight characteristics play an important role in the product, whether or not it becomes a healthy product. So, the products that do not have values for these characteristics are removed. 
Also, as we mentioned before, if the product $p$ in the \textit{variety} $Z$ which means $p \in Z$ and \textit{brand attribute} $Ba$ include \textit{variety} $Z$,  $Z \subset Ba$, the HTH-BD approach recommends $p_a$ alternative products within a \textit{variety} $Z$, then the products $p$ are analyzed within the \textit{variety} $Z$, and $Z$ that contain less than four products are removed, which is the first quarter of the value of the products for each \textit{variety}, so it becomes the minimum \textit{variety} that contains 4 products and the maximum number of product per \textit{variety} is 203, the median is 13 products and the mean contains about 24.3 products for the one \textit{variety}, this is after analyzing the products $p$ of HTH-BD approach, shown in~\autoref{tab:pr-variety}. 

After that, we  check the values of nutrition tables characteristics that have the same unit of measurement such as (Grams, \%, etc.). It turns out that the nutrition tables characteristics are measured in grams except the percentage of good fats $gf$ and dietary fiber $df$, and each of them is measured in percentage. They are converted to grams~\cite{percentagetogram, smith1972food} using $\scriptstyle  gf = \;$ ${\scriptstyle (gf / 100) * f}$.
Also, converted the dietary fiber variable, $\scriptstyle  df = \;$ ${\scriptstyle (df / 100) * Carbs}$.

This approach has used some nutrition books and nutrition experts~\cite{flynn2020good,egnell2020objective,dreano2020performance,jamieson2020food} to arrange the nutritional table features used in this approach. The result of this arrangement was (\textit{protein}, then \textit{good fats}, then \textit{dietary fiber}, then \textit{salt}, then \textit{sugars}, then \textit{carbohydrates}, then \textit{saturated fat}, and finally \textit{fat}).
In our research, an additional weight value was added to each nutrition table feature to help us arrange the product alternative.

Let $\vec{h}$ be a sort of nutrition table features list with weight value of each nutrition table feature. Each element contains a vector $\vec{h}_i$. Likewise, $\vec{h_i} = (pn_i*100 + gf_i*200 + df_i*300 + sa_i*400 + s_i*500 + Carbs_i*600 + sf_i*700 + f_i*800)$. It stores \textit{protein} $(pn)$ , \textit{good fats} $(gf)$, \textit{dietary fiber} $(df)$, \textit{salt} $(sa)$, \textit{sugars} $(s)$, \textit{carbohydrates} $(Carbs)$, \textit{saturated fat} $(sf)$, and \textit{fat} $(f)$ nutrition features about the products. 

Let $\vec{Nt}$ be a $|Z|$-dimensional vector. Here, $\vec{Nt} = (\vec{h}_1 , \dots, \vec{h}_|Z|)$ where $Nt_i = (\vec{h_i})$. 
Let $d_h(q_i,q_j)$, this denotes the following similarity measure according to their nutrition table of the product $q_i$ with respect to the product $q_j$~\autoref{eq:helth_dis}, the similarity calculated based on the output of PRO-COM ~\autoref{eq:price_3}.

\begin{equation}
\label{eq:helth_dis}
d_h(q_i,q_j) = Nt_{q_j}/d_m(p,q_i)
\end{equation}

Being the product unavailable, the alternative product $p_a|_{h}$ is selected according to that measure. The less value in the alternative product becomes a healthy product for the user:
    \begin{equation}
    \label{eq:alternative-product}
    p_a|_{h} = p_{a_i}|_{h} \;\; / \;\; d_h(p,p_{a_i}|_{h}) = \min_{\substack{\: \forall q_{i} \in \vec{p}_a|_b\\ p \neq q_i}} \: \{d_h(p,q_i)\} \;
    \end{equation}

If there is more than one $p_a|_h$ value, a $u_h$-dimensional vector ${\vec{p}_a}|_h = (p_{a_1}|_h, \dots, p_{a_{u_h}}|_h)$ is created being $u_h$ the number of alternatives.

\subsection{RS-NN: Package-based Approach}
\label{sec:thirdapp_ml}

The Package-based (PK-BD) approach is considered to include all the approaches together as it depends on the PRO-COM and HTH-BD approaches. The algorithm was developed based on the result of the HTH-BD approach. First, products that do not contain values for the three variables, which are product \textit{size}, \textit{units of measure}, and \textit{servings} are removed, and these are the variables on which this approach depends. 
Second, as mentioned above, the product $p$ and alternative products $pa$ must be within a \textit{variety} $Z$, and within a \textit{brand attribute} $Ba$, so the quantity of products within the varieties is analyzed so that the varieties containing less than the first quarter value are removed from the number of products within each $Z$ and its value is 4. Therefore, in the PK-BD approach as shown in~\autoref{tab:pr-variety}, the minimum product per \textit{variety} is four products, and maximum of the product per \textit{variety} is 203 products, and the median number of products is 13, and the average becomes 24.3 products. 
The algorithm is based on arranging alternative products $p_a$ based on the \textit{servings} $Sg$ value of the product $p$.
Let $\vec{Sg}$ be the vector that contains the \textit{servings} of the products in $Z$ ($|Z|$-dimensional vector). Hence, $\vec{Sg} = (Sg_1,\dots,Sg_{|Z|})$. The value of \textit{servings} $Sg$ in the product $q_i$ is compared to the alternative product $q_j$ , and there are two possibilities: namely that the product $q_i$ has \textit{servings} $Sg$ value greater than the \textit{servings}  $Sg$ value of the alternative product $q_j$, or vice versa. 
Let $d_{se_i} = d_{se}(q_i,q_j)$. This denotes the following similarity measure of the product $q_i$ with respect to the product  as shown in~\autoref{eq:size_step1}.

\begin{equation}
        \label{eq:size_step1}
        d_{se}(q_i,q_j) = \Bigg\{
        \begin{tabular}{l}
        ${\scriptstyle (Sg[q_i]\; /\; Sg[q_j])\; / \;\scriptstyle d_h(q_i,q_j)\;,}$ ${\scriptstyle \textrm{if} \;}$ ${\scriptstyle Sg[q_j]}$ ${\scriptstyle >}$ ${\scriptstyle Sg[q_i]}$\\
        ${\scriptstyle (Sg[q_j]\; /\; Sg[q_i]) \; / \;\scriptstyle d_h(q_i,q_j)\;,}$
        ${\scriptstyle \textrm{if} \;}$ ${\scriptstyle Sg[q_j]}$ ${\scriptstyle \leq}$ ${\scriptstyle Sg[q_i]}$\\
        \end{tabular}
\end{equation}

The distance is additionally applied to the products 
in order to select the alternative product $p_a$ as shown in~\autoref{eq:size_3}. 
There is the output more than one alternatives product $p_a|_s$ value, a $j$-dimensional vector $\vec{p}_a|_s = (p_{a_1}|_s, \dots, p_{a_j}|_s)$ is created being $j$ the number of alternatives.
    \begin{equation}
    \label{eq:size_3}
    p_a|_{s} = {p_{a_i}}|_{s} \;\; / \;\; d_{se}(p,{p_{a_i}}|_{s}) = \max_{\substack{\: \forall q_{i} \in \vec{{p}_a}|_{h}\\ p \neq q_i}} \: \{d_{se}(p,q_i)\} \;
    \end{equation}
    
The arrangement of alternative products $p_a|_s$ is based on the closest similarity ratio to the product $p$, taking into account the value of \textit{servings} $Se$, which is greater or less in proportion to the value of the \textit{servings} $Se$ of the product $p$.

 \section{Experimental Evaluation} 
\label{sec:results} 

In order to evaluate the effectiveness and performance of our recommender system, which is exclusively based on the product characteristics, we have used the following the following hardware and software equipment. We have selected the Python language to implement the different recommender system approaches. Our system uses the Window 10 operating system, and a hardware with the following specifications: Intel(R) Core(TM) i7-5500U, CPU (2.4 GHz), RAM (16 GB), and Storage (1 TB). The response time is really low, being approximately 4 s. to recommend alternative products of each desired product. In order to perform these tests, we randomly selected the products from MDD-DS for each approach. The alternative products given by the different RS approaches are displayed and stored in a report, for the users and experts to check the results. Finally, and in order to deployed the survey, we have decided to conduct a web survey using Python~\cite{swamynathan2019mastering} and Django~\cite{doan2009developer, stiglerpractical} on Amazon Elastic Compute Cloud (Amazon EC2), with the following specifications: vCPU (8), Memory (32 GiB), Network Burst Bandwidth (Up to 5 Gbps).

\begin{figure}[H] \centering 
\includegraphics[scale = 0.50]{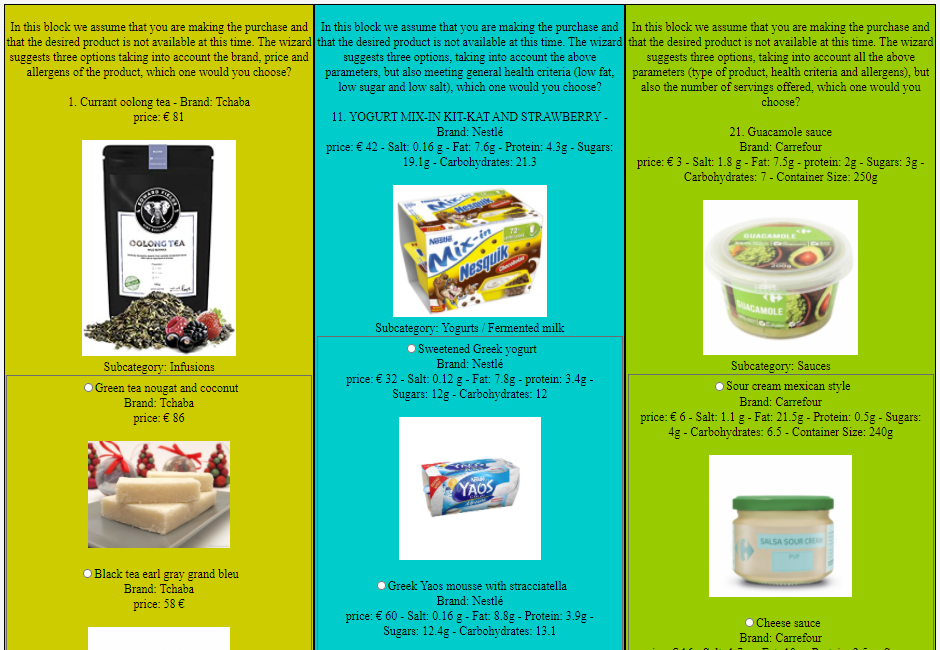} \caption{\label{fig:survey}A snapshot of the web survey.}  
\end{figure}

The evaluations are presented through a survey that includes three approaches, which it is answered by users and experts. Each survey comprises 30 questions and each group contains 10 questions as shown in~\autoref{fig:survey}. 
Each question includes a product $p$ and three alternative products $p_a$, with alternative products being the first three products with the closest value to the product. The survey depends on the situation in which the person is shopping and the product has not been found, so the user chooses between three alternative products according to each approach.

The experimental evaluation are divided into four subsections, which are the evaluations obtained from the user (\autoref{sec:resultsnon-ML user survey}), in \autoref{sec:results-ML Expert}, the evaluations obtained by the expert, in \autoref{sec:results-ML user survey}, the evaluations obtained by the user based on the results of the expert, and finally, evaluation and discussion which is a comparison between user surveys (\autoref{sec:Evaluation}).

\subsection{RS-CF User Survey}
\label{sec:resultsnon-ML user survey} 
The survey is divided into three blocks. 
Therefore, the first block is considered expressing the PRO-COM approach ~(\autoref{sec:procom}), the second block is dedicated to the PK-BD approach~(\autoref{sec:pkg_app}), and the last block is also performed to evaluate the HTH-BD approach~(\autoref{sec:helthapp}).

Survey results were calculated using mean squared error (MSE) after 65 people had responses. That said, once the products are sorted, there may be more than one substitute. Therefore, a link may occur between the highest-rated products recommended in the order given by the various approaches of the $RS$, so that the MSE~\cite{torabi2010mean} has been calculated for each approach taking into account the following three groups:
\begin{itemize} 
\item Group 1: All the questions answered by the users are considered.
\item Group 2: The questions having untied answers and the questions in which just the top-2 choices are tied are considered.
\item Group 3: Just questions with untied answers are considered.
\end{itemize}
The formula used to calculate the MSE is the following:

\begin{equation}
    \label{eq:mse}
    MSE = \frac{1}{n}\sum_{i = 1}^{n} (\hat{Y_i} - Y_i)^{2}
\end{equation}

Where the value $\hat{Y_i}$ is the value of the answer chosen by the user and $Y_{i}$ is the top-1 product, having always a value of 1. The value of $\hat{Y_i}$ would be $(1; 0.5; 0)$ if the user chose the first, second or third product of the survey, respectively. The values would be $(1; 1; 0.5)$ if there is a tie between the top-2 products and it would be $(1; 1; 1)$ if the tie happened between all the products. The results are shown in~\autoref{tab:result_mse1}, taking into account the three groups.

\begin{table}[htb] \centering
\caption{The MSE considering the three approaches as well as the different groups of products tested.}
\begin{tabular}{cp{40pt}p{40pt}p{40pt}p{40pt}} \hline
\multicolumn{1}{c}{\textbf{Approaches}}&\textbf{Group 1} &\textbf{Group 2} &\textbf{Group 3}\\ \hline
PRO-COM & 0.13885& 0.21539 &0.26187  \\ \hline
PK-BD & 0.22423 &0.25304 &0.26695  \\\hline
HTH-BD  &0.33731 & 0.36002 &0.36002 \\\hline
\end{tabular}  
\label{tab:result_mse1} 
\end{table}

Accuracy (ACC) was also calculated for the result (only for Group 3) as shown in~\autoref{eq:accuracy}. 

\begin{equation}
    \label{eq:accuracy}
    ACC = \frac{\sum_{i = 1}^{n} ((x_i= 1) \lor (x_i= 2)) }{n}
\end{equation}

where the value $n$ is the number of questions in group 3 and $x_i$ is the answer chosen by the user of group 3. It includes the answers in which the choice of the first or second product will be declared as positive while the third product will be declared negative. The result is shown in~\autoref{tab:result_accuracy_g3}.

\begin{table}[htb] \centering
\caption{The accuracy considering the group 3.}
\begin{tabular}{cp{40pt}p{40pt}} \hline
\multicolumn{1}{c}{\textbf{Approaches}} &\textbf{Group 3}\\ \hline
PRO-COM &81.33\%  \\ \hline
PK-BD & 79.28\%  \\\hline
HTH-BD & 70.77\%   \\\hline
\end{tabular}  
\label{tab:result_accuracy_g3} 
\end{table}

\subsection{RS-NN Expert Survey}
\label{sec:results-ML Expert} 
The company provided experts to evaluate the three approaches in the recommendation system, and expert opinions are important in evaluating the recommendation system for several reasons, the most important of which is that the experts fully know the products and also know the alternative products, so they can easily give their opinion that the recommendation system recommends suitable alternative products or not. Four surveys were sent of each approach to the experts, the surveys are the result of the techniques used; are (Cosine, Jaccard, Euclidean and Manhattan similarity) those mentioned in~\autoref{sec:simi_techq}. In each survey, the expert must answer three questions, namely: (1) Would you select any of these 3 options (alternative products)? (Yes/no). (2) If yes, select which one? (for example, 3). (3) Elaborate a raking to order the options (from the most similar product to the less similar product). Example: 3, 1, 2.

\begin{figure}[ht!]
    \includegraphics[width=0.3\textwidth]{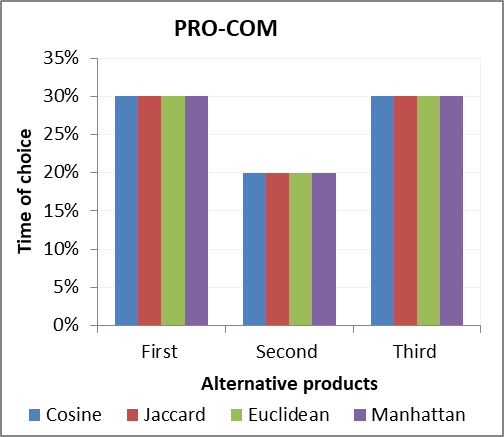}\hfill
    \includegraphics[width=0.3\textwidth]{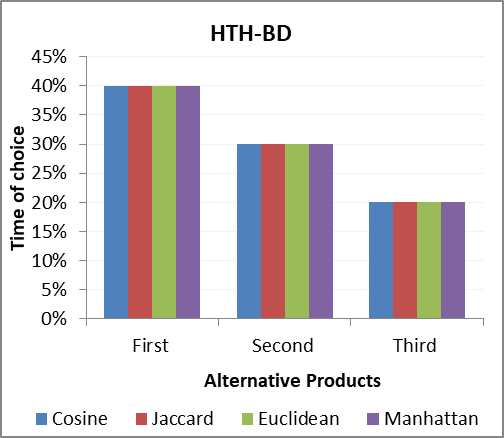}\hfill
    \includegraphics[width=0.3\textwidth]{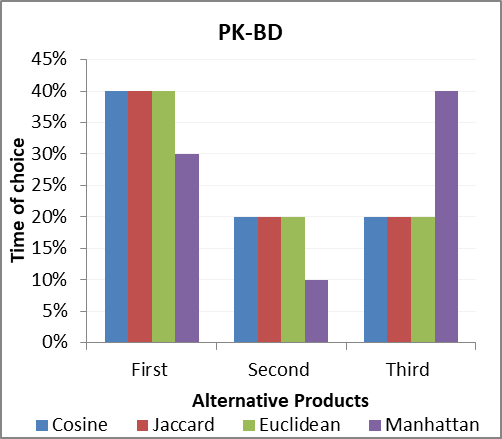}
\caption{The number of times the alternative products was chosen}
\label{fig:times}
\end{figure}

\autoref{fig:times} shows the results of the surveys indicating how many times the alternative product, be it the first, second or third, was chosen for each technique and also for each approach. The results show that the first approach has 80\% of the questions that have suitable alternatives. 
The expert reported that the second approach recommends that 90\% of the questions have suitable alternative products. The expert also stated that the third approach also recommends 80\% suitable alternative products.

Therefore, a user survey was created based on the opinions of the experts. Cosine similarity was chosen for all three approaches.

\subsection{RS-NN User Survey}
\label{sec:results-ML user survey}
This survey was built after taking the result from the expert, and this survey was very similar to the first survey we did, but augmented with clear images to make it easier for the user to quickly get to know the product and choose between alternative products, it is easier than. 
This survey is considered including three blocks. The first block expresses the PRO-COM approach~(\autoref{sec:firstapp_ml}), the second block is dedicated to the HTH-BD approach~(\autoref{sec:secondapp_ml}), and the last block is also implemented to evaluate the PK-BD approach~(\autoref{sec:thirdapp_ml}).

As shown in~\autoref{tab:result_mse2}, the survey results were also calculated using MSE as we did in the first survey after receiving 65 responses from users. The same groups that were used before were used to compare the results between the two investigations.

\begin{table}[htb] \centering
\caption{The MSE considering the three approaches, the three groups of products evaluated (Second survey).}
\begin{tabular}{cp{40pt}p{40pt}p{40pt}p{40pt}} \hline
\multicolumn{1}{c}{\textbf{Approaches}}&\textbf{Group 1} &\textbf{Group 2} &\textbf{Group 3}\\ \hline
PRO-COM & 0.213846 & 0.229372 & 0.235824   \\ \hline
HTH-BD & 0.210769 & 0.210769 & 0.210769 \\ \hline
PK-BD & 0.180769 & 0.188 & 0.188 \\ \hline
\end{tabular}  
\label{tab:result_mse2} 
\end{table}

Accuracy (ACC) for the result of Group 3 was also calculated as shown in~\autoref{tab:result_accuracy2} using~\autoref{eq:accuracy} as calculated in the first survey.

\begin{table}[htb] \centering
\caption{The accuracy of user survey using ML (group 3).}
\begin{tabular}{cp{40pt}p{40pt}} \hline
\textbf{Approaches} &\textbf{Group 3}\\ \hline
PRO-COM & 82.13058\%  \\ \hline
HTH-BD & 83.38461\% \\ \hline
PK-BD &  87.68\% \\ \hline
\end{tabular}  
\label{tab:result_accuracy2} 
\end{table}

\subsection{Evaluation and Discussion}
\label{sec:Evaluation}

Performing offline experiments by using a pre-collected data set to let users choose or rate items is a usual way to estimate the performance of recommender systems, such as prediction accuracy~\cite{shani2011evaluating}. In this case, the dataset is usually divided into (i) a training sample to build the model based on the user rating and (ii) a test sample to calculate the measurement parameters such as accuracy, precision, recall and f-score. Since our recommender system is uniquely based on the characteristics of the products and we are not considering the customer profile, this kind of offline experiment is not provided for our evaluation purposes.

However, we have decided to opt by a most direct evaluation based on the feedback from two important sectors: customers (users) and experts (workers in the food retail sector). For this, we created a large-scale experiment on a prototype through a user survey, that is, an online experiment. The results is the direct feedback and opinion of the performance of the recommender system according to the users' perspective. Consequently, the feedback obtained would depend on a variety of factors, such as the user's intent (for example: how specific are their information needs), the user's context (for example: what items are they already familiar with, in addition, how much they trust the system) and the interface which the recommendations are presented through. This is a more realistic scenario and it will give as strong evidences about the recommender system's results: that is, if the suggested product is on the user would buy instead of the required one or not, obtaining, therefore, a good value for the accuracy.

Since we do not have information about the costumers (profiling, interactions...), we have worked on the whole data without dividing the data set. In order to save time and optimize operational performance when recommending alternative products, we took the following steps. We filter and pre-process the data set by two methods (BOW, Doc2Vec) for each approach based on the desired product characteristics, such as (variety, size, allergen, etc.). In RS-CF, we compare the desired product with the rest of the data and order the alternative based on the similarity ratio. In RS-NN, we build the model for the desired product using the neural network and then classify the alternative based on the similarity ratio.

\begin{figure}[htp] \centering \includegraphics[scale = 0.50]{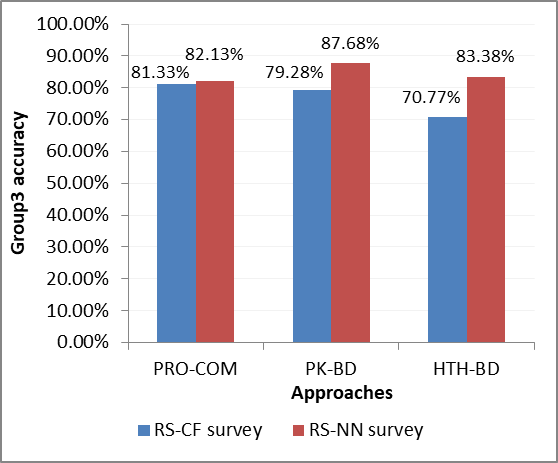} \caption{Comparative study using accuracy considering the group 3.   \label{fig:accuracy_group3}} \end{figure}

\begin{figure}[ht!]
    \includegraphics[width=0.3\textwidth]{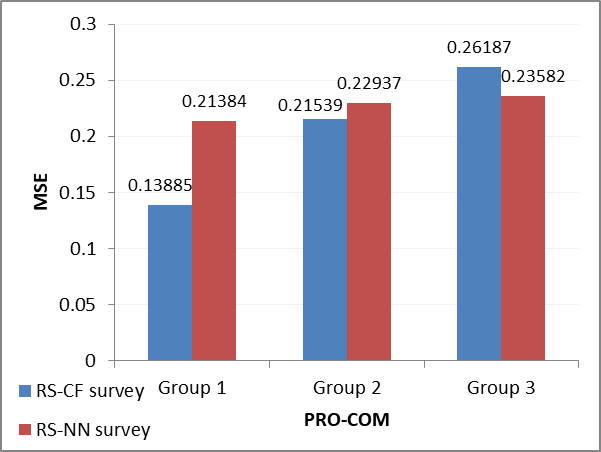}\hfill
    \includegraphics[width=0.3\textwidth]{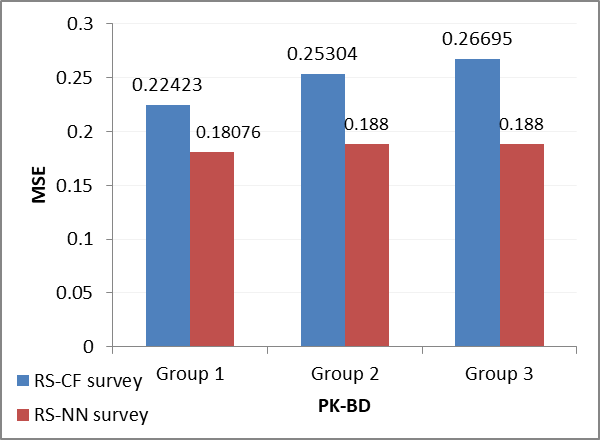}\hfill
    \includegraphics[width=0.3\textwidth]{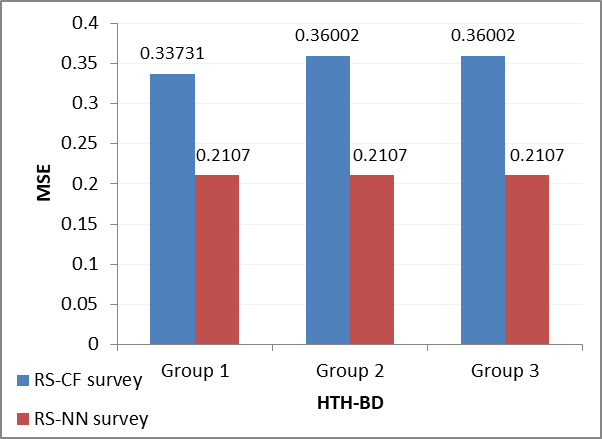}
\caption{The MSE of three approaches}
\label{fig:MSE}
\end{figure}

Surveys evaluated the recommendation system, which is RS-CF user survey and RS-NN user survey, where comparison was made between them, as shown in~\autoref{fig:accuracy_group3}, which shows 
the difference between the accuracy results of the two surveys considering group 3.
The results showed that the RS-NN user survey performed better in all three approaches.
Also,~\autoref{fig:MSE} shows the difference between the MSE results as it showed that the RS-CF user survey results are the best for the PRO-COM approach for the first and second groups, but the RS-NN user survey is the best for the third group. Also, the RS-NN user survey is the best for both approaches: PK-BD and HTH-BD.
The comparatives prove that using the neural network-based completely alters the results, and taking price and brand into account was something that users wanted. Also, using more nutrition table features gives better results. It also proved that a PK-BD approach based on the HTH-BD approach is far better than relying solely on a PRO-COM approach.

Finally, we evaluated the RS with multi-criteria through a user survey using MSE to calculate the average error for the responses of the users of three groups, which is the main evaluation of the users' responses. We also use accuracy to evaluate the responses of users in group 3 only because for the other two groups it approximately resulted a $ 100\%$.

\section{Conclusions}
\label{sec:conclusions}

The recommendation idea is to implement some approaches that help the user get the right product. The approaches are made based on the user's interest. For example, suppose the user is interested in a specific size product or product that does not contain an allergen, it is not available in stock. In that case, the RS recommends an similar product with these specifications without referring to the user's file; recommended depending only on the user's choice. Besides, the recommendation system can recommend an alternative health product to the user. In this paper, to build a recommendation systems, we used item-based collaborative filtering (RS-CF) and BOW to represent the dataset as a vector. To build an RS-CF that caters to the largest number of users. We created three approaches, which are product composition (PRO-COM), package-based (PK-BD), and the healthy-based approach (HTH-BD). Essentially, PRO-COM works to obtain a similar product based on the product's component. Whereas, PK-BD approach takes into consideration PRO-COM and adds product size to obtain a similar product. Finally, the HTH-BD approach obtains a similar product taking PRO-COM and allergen information into account, then which an equation is made consisting of the features of the nutrition table. The user then evaluates these approaches through the survey.

After that, we refine the recommendation to suit the company's requirements. Optimization the RS-NN model is done using the neural network-based as a representation dataset and create a model using Doc2Vec. RS-NN tries to improve the approaches also by adding some considerations (like allergen feature as apre-condition for all approaches, more features about nutrition table and brand type, brand attribute and price) and rearranging approaches to PRO-COM, HTH-BD, then PK-BD.
 
A survey of experts and users was conducted to assess RS-NN. Then we collected the result that we compared between the models (RS-CF, RS-NN), the comparatives prove that using the neural network-based completely alters the results.

For future work, develop this research, especially the health approach, so that this approach is based on the user's profile and not just on the product's components. 
For example, when creating a user profile such as age, chronic disease, prominent diet, etc. This profile helps the recommendation system to recommend a suitable alternative healthy product for the user.

\section{Acknowledgments}
The authors would like to thank the European Regional Development Fund (ERDF) and the Galician Regional 7 Government, under the agreement for funding the Atlantic Research Center for Information and Communication Technologies (AtlanTTIC), and the Spanish Ministry of Economy and Competitiveness, under the National Science Program (TEC2017-84197-C4-2-R).


\bibliographystyle{ieeetr}
\bibliography{2020-Sensors-Manar-Grocery}

\end{document}